\documentclass[fleqn,usenatbib]{mnras}

\usepackage[T1]{fontenc}
\usepackage{ae,aecompl}
\usepackage{amstext}
\usepackage{url}
\usepackage[]{natbib,amsmath,amssymb,times,bm}
\bibpunct{(}{)}{;}{a}{}{,}
\usepackage{tabularx,amsmath,amssymb,hyperref}
\usepackage{mathrsfs} 
\usepackage{graphicx,epsfig,color,latexsym}
\usepackage{array}
\usepackage{booktabs}
\usepackage{multirow}
\usepackage{changepage}
\usepackage{xcolor}
\usepackage{ulem}

\newcommand{\bea}{\begin{eqnarray}}
\newcommand{\eea}{\end{eqnarray}}
\newcommand{\be}{\begin{equation}}
\newcommand{\ee}{\end{equation}}

\renewcommand{\exp}{\mathrm{exp}}

\usepackage{xcolor}

\title[$f(T)$ lensing]{Exploring \texorpdfstring{\bm{$f(T)\;$}}. Gravity via strongly lensed fast radio bursts}

\author[Jiang et al.]%
       {Xinyue Jiang$^1$\thanks{xinyue.jiang@mail.ynu.edu.cn}, Xin Ren$^{4,2,3}$, Zhao Li$^{2,3}$, Yi-Fu Cai$^{2,3}$, Xinzhong Er$^1$\thanks{phioen@163.com}\\       
$^1$ South-Western Institute for Astronomy Research, Yunnan University, Kunming, P.R.China\\
$^2$ CAS Key Laboratory for Researches in Galaxies and 
Cosmology/Department of Astronomy, School of Astronomy and Space Science, \\
University of Science and Technology of China, 96 Jinzhai Road, Hefei, Anhui 230026, China\\
$^3$ Deep Space Exploration Laboratory/School of  Physical Sciences, 
University of Science and Technology of China, Hefei, Anhui 230026, China\\
$^4$ Department of Physics, Tokyo Institute of Technology 2-12-1 Ookayama, Meguro-ku, Tokyo 152-8551, Japan
}

\date{Accepted XXX. Received YYY; in original form ZZZ}

\begin{document}
\maketitle

\begin{abstract}
This study aims to investigate the strong gravitational lensing effects in $f(T)$ gravity. We present the theoretical analytic expressions for the lensing effects in $f(T)$ gravity, including deflection angle, magnification, and time delay. On this basis, we also take the plasma lensing effect into consideration. We compare the lensing effects between the General Relativity in a vacuum environment and the $f(T)$ gravity in a plasma environment. From a strongly lensed fast radio burst, the results indicate that in a plasma environment, General Relativity and $f(T)$ gravity can generate indistinguishable image positions, but the magnification and time delay on these positions are significantly different, which can be distinguished by current facilities in principle. Therefore, the discrepancies between observational results and theoretical expectations can serve as clues for a modified gravity theory and provide constraints on $f(T)$ gravity.

\end{abstract}


\vspace{1.0\baselineskip}

\section{Introduction}
\label{sect:intro}

The acceleration of the cosmic expansion is a pivotal discovery in modern cosmology \cite[]{Riess1998ObservationalEF,Perlmutter1998MeasurementsO}. To explain the accelerating expansion of the universe, two scenarios have been proposed, one is to introduce the cosmological constant or "dark energy" under the framework of General Relativity (GR), postulating the existence of new, unknown forms of energy driving the universe's expansion \cite[e.g.][]{Clowe2006ADE,Frieman2008DarkEA,Caldwell1997CosmologicalIO}. The other explanation emerges from modified gravity theories \cite[e.g.][]{Clifton2011ModifiedGA}, which usually introduce modifications to the Einstein-Hilbert action through the addition of extra terms \cite[e.g.][]{Nojiri2006MODIFIEDF,DeFelice2010fRT}. Similarly, the corresponding Lagrangian can be extended in various ways from the Teleparallel Equivalent of General Relativity (TEGR), which uses torsion instead of curvature to describe gravity \cite[]{Unzicker2005TranslationOE,Cai2011MatterBC,Hohmann2022TeleparallelG,Maluf2013TheTE,Krk2018TeleparallelTO,Bahamonde2021TeleparallelGF}. It has been demonstrated that these modified torsional gravity theories, particularly the $f(T)$ theory, exhibit reliable performance in cosmology. It can explain not only the early inflation but also the current acceleration of the expansion of the universe \cite[e.g.][]{Cai2015fTTG, Bengochea2008DarkTA, Zheng2010GrowthFI, Bamba2010EquationOS, Cai2011MatterBC, CAPOZZIELLO2011167, Bamba2013ConformalSA, Farrugia2016StabilityOT, Yan2019InterpretingCT, Wang2020CanF, Li2018TheEF, Ren2021DatadrivenRO, Ren2022GaussianPA, Zhao2022QuasinormalMO,  Bhmer2011ExistenceOR}.

For the $f(T)$ power law model, represented as $f(T)=T+\bm{\mathring{\alpha}}T^n$, extensive research has been conducted to investigate the constraints on the spherically symmetric solutions of $f(T)$ gravity through solar-system experiments \cite[e.g.][]{Iorio2012SolarSC,Iorio2016ConstrainingTS,Xie2013fG}. For example, in \cite{Ruggiero2016LightBI}, an upper limit for the parameter $\bm{\mathring{\alpha}}$ is estimated, $5 \times 10^{-1}\mathrm{m}^2$, giving a relatively stringent numerical constraint. To further determine the range of the $\bm{\mathring{\alpha}}$, experiments beyond the solar system are necessary. The phenomenon of strong gravitational lensing offers an extension of the experimental scope from the solar system to the cosmological scale. 

The universe contains massive objects, such as galaxies and galaxy clusters, which possess strong gravitational potential \cite[]{Kravtsov2012FormationOG}. When passing near massive objects, the light emitted from distant sources experiences deflection, resulting in the formation of distorted images of the source, a phenomenon known as gravitational lensing \cite[]{Schneider1992GravitationalLA, Blandford1992CosmologicalAO,Refsdal1964TheGL, Surdej1993GravitationalLI}. The magnification of these images is a crucial aspect of gravitational lensing, it provides us with the opportunity to probe distant objects \cite[e.g.][]{Kelly2014MultipleIO}, and study the matter distribution of the lens \cite[e.g.][]{Blandford1986FermatsPC, Keeton2001ACO}. Another remarkable feature of gravitational lensing is the time delay effect between the multiple images of a single source. The time delay is sensitive to both the geometry of the universe and mass distribution of the lensing object \cite[e.g.][]{Suyu2012TWOAT, Bozza2003TimeDI, Oguri2001StrongGL, Suyu2012AccurateCF, Treu2022StrongLT, Gurkan2010AMF}. Consequently, the gravitational lensing effect has proven to be a powerful tool in astrophysics and cosmology, enabling researchers to probe the properties of both the lensing objects and the background sources, and also testing cosmological theories \cite[e.g.][]{Treu2010StrongLB, Hildebrandt2016KiDS450CP, Zhang2007ProbingGA}.

Gravitational lensing can be also used to constrain and validate different modified gravity theories \cite[e.g.][]{Nazari2022LightBA, Bahamonde2020SolarST, Ren2021DeflectionAA, Kuang2022ConstrainingAM, Yang2018NewPO, Narikawa2013TestingGS, Schmidt2008WeakLP}. Given the high precision achievable in gravitational lensing time delay measurements \cite[e.g.][]{Courbin2004COSMOGRAILTC, Chen2016SHARPI, Tewes2012COSMOGRAILTC}, we specifically focus on comparing the $f(T)$ gravity with General Relativity by lensing time delay. Therefore, rapid transient sources are necessary for background source \cite[][]{Oguri2019StrongGL,Liao2022StronglyLT}. Fast Radio Burst (FRB) has served as a powerful probe for studying theories related to cosmic accelerating expansion \cite[e.g.][]{Zitrin2018ObservingCP,Oguri2019StrongGL,Abadi2021ProbingGS}.
With the advantage of extremely bright pulses ($\sim 50 \text{mJy} - 100 \text{Jy}$) and short duration ($\sim$ milli-seconds) \cite[] {Spitler2016ARF, Lorimer2007ABM, Marcote2017TheRF, Cordes2019FastRB}, the time delays induced by lensing can be precisely estimated from the observational data and provide tight constraints to cosmology \cite[e.g.][]{Wu2014ConstrainsOF,Abadi2021ProbingGS,Gao2022ProspectsOS,Wucknitz2020CosmologyWG,Muoz2016LensingOF}. The predicted number of events of lensed FRBs is reasonable in the future \cite[]{2023MNRAS.521.4024C}.

The emission of FRBs mainly falls in low-frequency radio bands. The radio signals from cosmic distances are refracted as they travel through cold plasma thus resulting in plasma lensing effect \cite[e.g.][]{Clegg1997TheGP,Tuntsov2015DYNAMICSM}. The plasma in the lens galaxy can induce such an effect on the signal of a lensed FRB. In this work, we address the lensed FRBs produced by $f(T)$ gravity in a plasma environment and discuss their differences from the gravitational lensing effects in GR. Additionally, we give predictions of the lensing effect generated by plasma in both GR and $f(T)$ for various $\bm{\mathring{\alpha}}$ values.

The outline of this paper is as follows. In Section.\,\ref{sect:lensing framework}, we give a brief introduction to the lensing framework. In Section.\,\ref{sect:lensing effect under $f(T)$ theory}, we derive the gravitational lensing effects under $f(T)$ gravity. In Section.\,\ref{sect:plasma lensing}, the plasma lensing effects are included as well for strongly lensed radio sources. We compare the $f(T)$ gravity with GR by simulating an example of lensed FRB. Finally, Section.\,\ref{sect:summary} gives the discussion and summary.

\section{Lensing framework}
\label{sect:lensing framework}

We outline the basic formulae of the gravitational lensing effect in this section, one can find more details in \cite[]{Meneghetti2021IntroductionTG}. 
Thin lens approximation and weak field approximation are adopted in this work, i.e. the impact parameter is much larger than the Schwarzschild radius of the lens \cite[]{Narayan1996LecturesOG,Blandford1992CosmologicalAO}. The difference between the GR and $f(T)$ becomes significant in the case of a strong field, i.e., a small impact parameter. Such kinds of observations at the moment can be achieved only near the black hole, e.g., the shadow of black hole \cite[e.g.][]{Cunha2018ShadowsAS,Johannsen2016TestingGR,Afrin2022TestsOL,Mizuno2018TheCA,Ayzenberg2018BlackHS,Psaltis2020GravitationalTB}. 
Thus we only consider the weak field cases in which the observation will be easier to perform \cite[e.g.][]{Abuter2018DetectionOT,Berti2015TestingGR,Ferreira2019CosmologicalTO,Hees2017TestingGR}. Here, we present the relevant coordinate definitions for lensing: we begin by considering an optical axis that is perpendicular to both the lens and source planes, passing through the observer. On the source plane, $\vec{\beta}$ represents the angular position of the source, while on the lens plane, $\vec{\theta}$ represents the apparent angular position of the received light and $\vec{\alpha}(\vec{\theta})$ denotes the deflection angle caused by the lensing effect. The effective gravitational lensing potential $\Psi(\vec{\theta})$ is defined as
\be
\Psi(\vec{\theta})\equiv \frac{D_{LS}}{D_{L} D_{S}} \frac{2}{c^{2}} \int \phi\, dZ,
\label{eq:lensing potential_gravity potential}
\ee
where $\phi$ is the gravitational potential, and $Z$ is the comoving angular diameter distance along the line-of-sight. $D_{LS}, D_{S}, D_{L}$ is, respectively, the angular diameter distance between lens and source, source and observer, lens and observer.
The gravitational lensing effects are fully determined by the surface mass density of the lens in the thin lens approximation. The dimensionless surface mass density $\kappa$, which also represents the lensing convergence and can be given by half of the Laplacian of $\Psi$,  
\be
\kappa(\vec{\theta})=\frac{1}{2} \triangle_{\theta} \Psi(\vec{\theta}).
\label{eq:kappa_lensing potential}
\ee
The gravitational lensing effects are determined by $\Psi$ (or $\kappa$) and then show up by the deflection of light. The deflection angle can be calculated from the gradient of lens potential 
%
\be
\vec{\alpha}(\vec{\theta})=\vec{\nabla}_{\theta} \Psi(\vec{\theta}).
\label{eq:alpah_gravity potential}
\ee
For simplicity, we will use dimensionless coordinates. The image position $\vec{x}$ obtained from the observation can be related to the deflection angle $\vec{\alpha}(\vec{x})$ and source position $\vec{y}$ through the lens equation:
\be
\vec{y} = \vec{x} - \vec{\alpha}(\vec{x}),
\label{eq:lens equation}
\ee
where $\vec{y} = \vec{\beta}/\theta_E$, $\vec{x} = {\vec{\theta}}/{\theta_E}$, and $\vec{\alpha}(\vec{x}) ={{\vec{\alpha}(\vec{\theta})}}/{\theta_E}$. The Einstein radius, denoted as $\theta_E$, is the radius of the Einstein ring, marking the boundary inside which strong lensing effects become noticeable. This radius reflects the total mass of the lensing object and influences the distances between the lensed images \cite[][]{Meneghetti2021IntroductionTG}.

The surface mass density $\kappa$ can be estimated from observation, such as image positions, distortions, and magnifications, but degeneracy exists in general \cite[][]{Schneider2014GeneralizedMG}. 
If the angular size of the source is small, the lensing image distortion can be described by the Jacobian matrix of the lens equation
\be
\mathcal{A} \equiv \frac{\partial \vec{y}}{\partial \vec{x}}=\left(\delta_{i j}-\frac{\partial \alpha_{i}(\vec{x})}{\partial x_{j}}\right).
\ee
The determinant of $\mathcal{A}$ gives the magnification of the image $\mu$ by
\be
\mu \equiv \frac{1}{\operatorname{det} \mathcal{A}}.
\ee
Along the curve where det$\mathcal{A}=0$, the theoretical magnification becomes infinity, and the curve is called the critical curve on the image plane. The corresponding curve on the source plane is caustic. The shear $\gamma$, which characterizes the stretching for an extended source, can be calculated from the second derivative of the lensing potential (e.g. $\partial^2 \Psi /{\partial x_i \partial x_j} \equiv \Psi_{i j}$),
\begin{flalign}
\gamma_{1}&=\frac{1}{2}\left(\Psi_{11}-\Psi_{22}\right), \\
\gamma_{2}&=\Psi_{12}.
\end{flalign}
%

\section{lensing effect under \texorpdfstring{\bm{$f(T)$}} .  gravity}
\label{sect:lensing effect under $f(T)$ theory}

In this section, we derive the gravitational lensing effects for the power-law model of $f(T)$ gravity, i.e. $f(T) = T+ \bm{\mathring{\alpha}}T^2$ \cite[e.g.][]{Ruggiero2016LightBI,Ruggiero2015WeakFieldSS}. In the point mass model, the deflection angle of $f(T)$ gravity is given by \cite{Chen2020NewTO}:
\be
\alpha_{f(T)}(\xi)=\frac{4 G m}{c^{2} \xi}+\frac{40 \pi \bm{\mathring{\alpha}} }{\xi^{2}} = \alpha_{GR} + \frac{40 \pi \bm{\mathring{\alpha}} }{\xi^{2}},
\label{eq:D_alpha}
\ee
where $\xi=D_L\theta$ ($\theta\equiv|\vec{\theta}|$) is the impact parameter, $m$ is the mass of the point lens and $\bm{\mathring{\alpha}}$ indicate the departure of $f(T)$ from GR. The point model is used for compact objects such as neutron stars or black holes. For $f(T)$ gravity theory, the constraint parameter $\bm{\mathring{\alpha}}$ is generally considered to be a constant independent of mass. Predictions of different values of the $\bm{\mathring{\alpha}}$ between the $f(T)$ gravity theory and GR for the deflection angle have been proposed, and compared with actual observational data \cite[e.g][]{Iliji2018CompactSI,DeBenedictis2016SphericallySV}. Similarly, we assume that $\bm{\mathring{\alpha}}$ is a constant independent of the lens mass in this work.

\subsection{SIS model}
\label{sect:SIS model under f(T) gravity}
The Singular Isothermal Sphere (SIS) model is commonly used to describe the dark matter halo profile of a lens galaxy, particularly for the elliptical galaxies \cite[][]{Binney2008GalacticDS,https://doi.org/10.1002/asna.2113140412, 1987gady.book.....B, Hurtado2013GravitationalLB}. 
In this model, the mass density is inversely proportional to the radius to the power of two, i.e., $\rho\propto r^{-2}$, with $r\equiv(\xi^{2}+Z^{2})^{1/2}$. Such a profile can give a constant velocity dispersion $\sigma_{v}$ of the dark matter "particle" of the galaxy halo. The surface mass density of an SIS halo is
\be
\Sigma (\xi) = {\sigma^{2}_{v}}/{2G\xi}.
\ee
The Einstein radius of an SIS halo in GR is 
\be
\theta_E =\frac{4 \pi \sigma^{2}_{v}}{c^{2}}\, \frac{D_{L S}}{D_{S}}.
\ee
In order to calculate the effective lensing potential, we start from the gravitational potential of $f(T)$. By integrating Eq.\,\ref{eq:D_alpha} and comparing with the deflection angle from GR, one can obtain the gravitational potential of $f(T)$:
\be
\phi_{f(T)}=\phi_{GR}-20 \frac{\bm{\mathring{\alpha}} c^{2}}{r^{2}}.
\ee
The convergence under $f(T)$ gravity can be calculated using Eq.\,\ref{eq:kappa_lensing potential}:
\be
\kappa_{f(T)}=\frac{4 \pi G}{c^{2}} \frac{D_{L} D_{L S}}{D_{S}}\left[\Sigma-\frac{10 \bm{\mathring{\alpha}} c^{2}}{G \xi^{3}}\right].
\ee
It can be simplified to 
\be
\kappa_{f(T)} 
= \frac{1}{2x}-\frac{A}{2x^{3}},\quad
{\rm with}\quad 
A = \frac{ 80 \bm{\mathring{\alpha}} \pi D_{LS}}{ \theta_E^3 D_S {D_L}^2} ,
\ee
where $x\equiv \theta/\theta_E$ is the dimensionless coordinate. The parameter $A$ absorbs the redshifts ($z_{S}, z_{L}$), the Einstein radius $\theta_E$, and the $f(T)$-related parameter $\bm{\mathring{\alpha}}$. In this work, we consider the following ranges: $z_{L} \in (0.15,1.5)$ and $z_{S} \in (0.6,6.0)$, which encompass the redshift range of strongly lensed quasars observed to date\footnote{https://research.ast.cam.ac.uk/lensedquasars/}. For $\theta_E$ we assume a typical value of $1 ''$. As for the parameter $\bm{\mathring{\alpha}}$ in the power-law model of $f(T)$ gravity, $f(T)$ will regress to GR gravity when $\bm{\mathring{\alpha}}=0$. Thus, its lower limit often left unspecified \cite[][]{Bahamonde2021TeleparallelGF}, while its upper limit varies widely \cite[e.g.][]{Iorio2015ConstrainingFG, Chen2020NewTO, Ruggiero2016LightBI}. In this analysis, we adopt an upper limit of recent observation $\bm{\mathring{\alpha}} = 0.33 $ pc$^{2}$ \cite[][]{Chen2020NewTO}, which gives $A=1$. In order to explore a large portion of the possible values, we adopt $A$ in this analysis spanning from $1$ to $10^{-15}$.

In terms of polar coordinate defined by $\vec{x}=x(\cos\varphi, \sin\varphi)$, the lensing potential can be found by solving the Poisson equation:
\be
\frac{\partial^{2} \Psi_{f(T)}}{\partial x^{2}}+\frac{1}{x} \frac{\partial \Psi_{f (T)}}{\partial x}+\frac{1}{x^{2}} \frac{\partial^{2} \Psi_{f(T)}}{\partial \varphi^{2}}=2 \kappa_{f(T)}=\frac{1}{x}-\frac{A}{x^{3}},
\ee
which is
\be
\Psi_{f(T)}=x-\frac{A}{x}.
\ee
%
%
%
Then the deflection angle can be written by
\be
\alpha(x)_{f(T)}=1+\frac{A}{x^{2}}.
\label{eq:sis-alpha}
\ee
%
%

The expression of the deflection angle in modified gravity is typically more complex than Eq.\,\ref{eq:sis-alpha}, many of them show non-single power law behaviours \cite[e.g.][]{Nzioki2010AGA, Wei2014BlackHS, Nzioki2010AGA, Campigotto2016StrongGL, 2023arXiv231010346S}. And most works focus on the point lens \cite[e.g.][]{Alhamzawi2016GravitationalLI, Panpanich2019ParticleMA, Alhamzawi2016GravitationalLB}. In \cite{Enander2013StrongLC}, the deflection angle in Hassan-Rosen bimetric gravity is expressed as $\alpha(x) = \alpha_{GR}(1+c^2 e^{-z_{E}})$, where $z_{E} = m D_{L} \theta_{E} \sqrt {\frac{(c+c^{-1})(\beta_1 + 2\beta_2 c + \beta_3 c^2)}{q}} $ is a specific parameter that depends on the Einstein radius and the distance of the lens, and other parameters in the Hassan-Rosen theory: $m$, $c$, $\beta_i$ and $q$. When $z_{E}$ varies from $50$ to $100$, the addition term of deflection angle ranges from $10^{-5}$ to $10^{-27}$ arcsec (in case of $\theta_{E} = 1 ''$). In \cite{2021EPJC...81..109S}, the deflection angle in $f(R)$ gravity express as $\alpha = 16 \pi G \rho \left(\frac{r_{out}}{\xi}\right)^2 \left(-\frac{1}{2} e^{-(\frac{\xi}{r_{out}})^2} + \frac{\sqrt{\pi}}{4} (\frac{r_{out}}{\xi})\text{erf}\left(\frac{\xi}{r_{out}}\right)\right)$, where $\xi = x \theta_{E} D_{L}$ still represents the impact parameter, $r_{out}$ is the cutoff radius of the dark matter halo where $\rho(r_{out})=0$. For a case $r_{out} = 5$ kpc, dark matter halo core density $\rho = 10^{-14}$  kg m$^{-3}$ in general \cite[see][]{2007A&A...464..921S, 2012PhRvD..86d4038W} and the same $z_{L}$ as we use, the value of the deflection angle varies around $10^{-5}$ arcsec (in case of $x=1, \theta_{E} = 1''$).
Unlike our result, which decreases with $x$, these modified deflection angles also show non-monotonous behaviours with $x$.

\begin{center}
\begin{figure*}  
  {\mbox{\includegraphics[width=3.4 in]{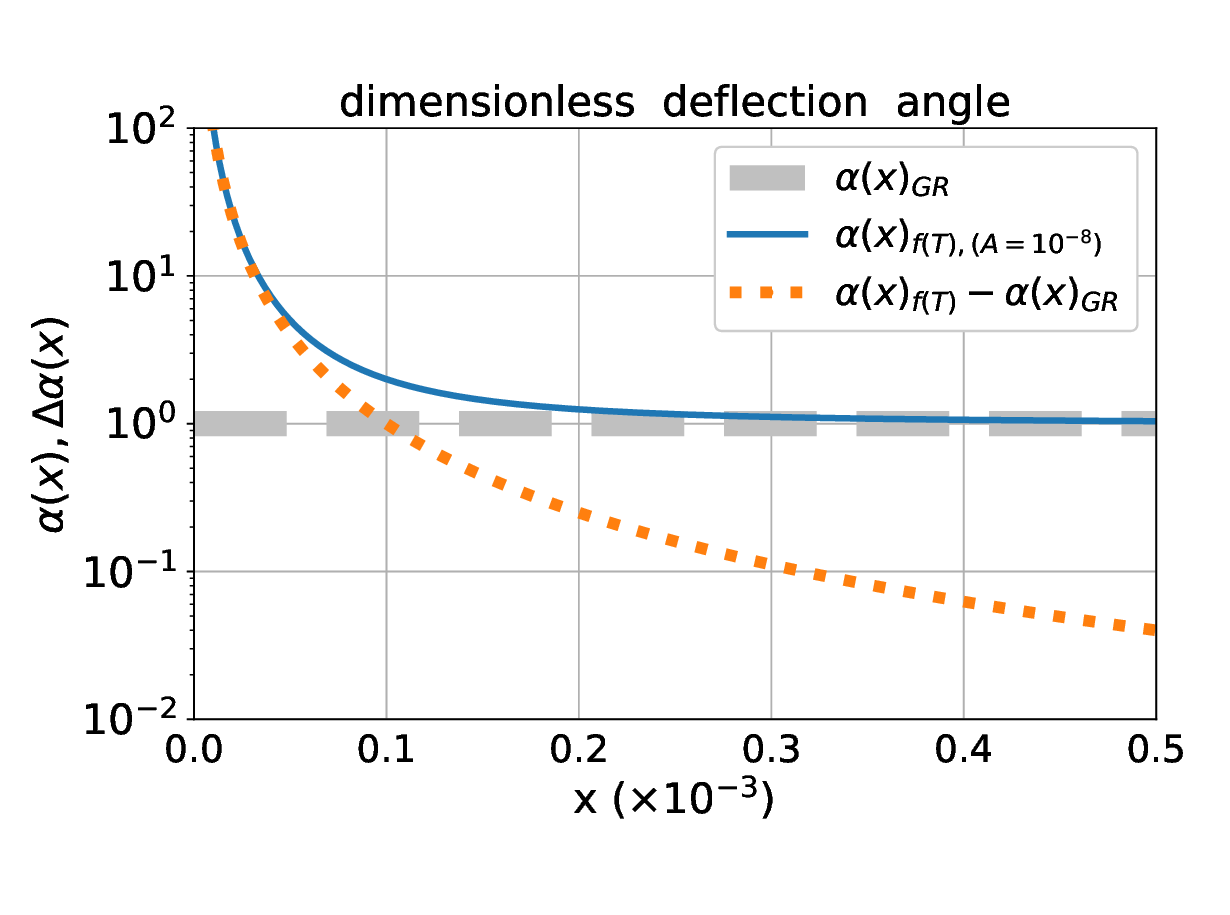}}{\includegraphics[width=3.4 in]{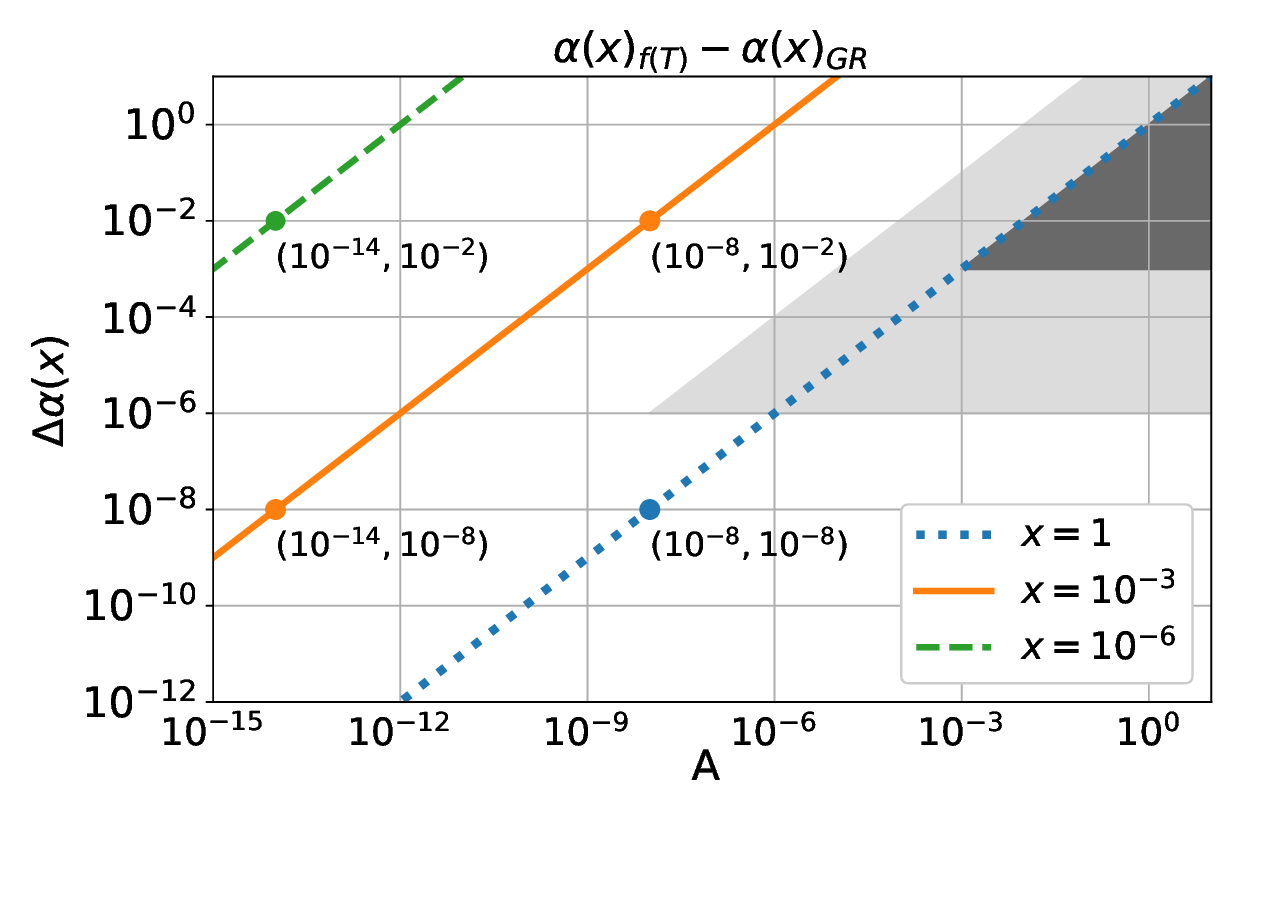}}}
  \caption{The left panel shows the dimensionless deflection angle in GR (gray dashed line), $f(T)$ gravity (blue solid line), and their difference (orange dotted line). The right panel shows $\alpha(x)_{f(T)}-\alpha(x)_{GR}$ with different values of $A$. Three image positions are shown with different colours. The coordinates of several points are presented for better comparison. The grey area shows the possible range that is distinguishable by the current telescope.}
\label{fig:SIS_alpha(x)}
\end{figure*}
\end{center}

Comparisons of dimensionless deflection angle under $f(T)$ and GR are shown in Fig.\,\ref{fig:SIS_alpha(x)}. Since the correction to the deflection angle is proportional to the parameter $A$, we choose a midpoint of the $A$ range, $10^{-8}$, to illustrate the differences in the deflection angle caused by GR and $f(T)$. In the left panel, we show the dimensionless deflection angle $\alpha_{GR}$, $\alpha_{f(T)}$, and $\Delta\alpha\equiv\alpha_{f(T)}-\alpha_{GR}$ as a function of image positions $x$. We find that the divergence of $f(T)$ only appears when the image forms close to the center of the lens. In the right panel, we show the difference in the deflection angle $\Delta\alpha$ as a function of the parameter $A$. Three colours are used to indicate three different image positions. We take the most accurate space-borne telescope, Gaia, as our reference, which can reach microarcsecond ($\mu$as) global astrometry\footnote{https://www.cosmos.esa.int/web/gaia/science-performance}. The radio observation by VLBI can achieve a milliarcsecond level of accuracy \cite[e.g.][]{2017isra.book.....T}. We adopt the limit accuracy of the image position by $10^{-6}$ arcsec and a typical Einstein radius $\theta_E\ = 1''$. The minimum separation between the lens and the images needs to be larger than the effective radius of the galaxy. We use an optimistic value of separation, 0.1 arcsec, which corresponds to a lens with effective radius of several kpc \cite[][]{Chen2018AssessingTE}. With that, we show the possible detection area for the parameter $A$ by the gray shadow in the right panel of Fig.\,\ref{fig:SIS_alpha(x)}. We also show a more conservative limit of image separation $1$ arcsec and image astrometry $10^{-3}$ arcsec by the darker region.

\begin{figure}
  \centering
  \includegraphics[width=3.3 in]{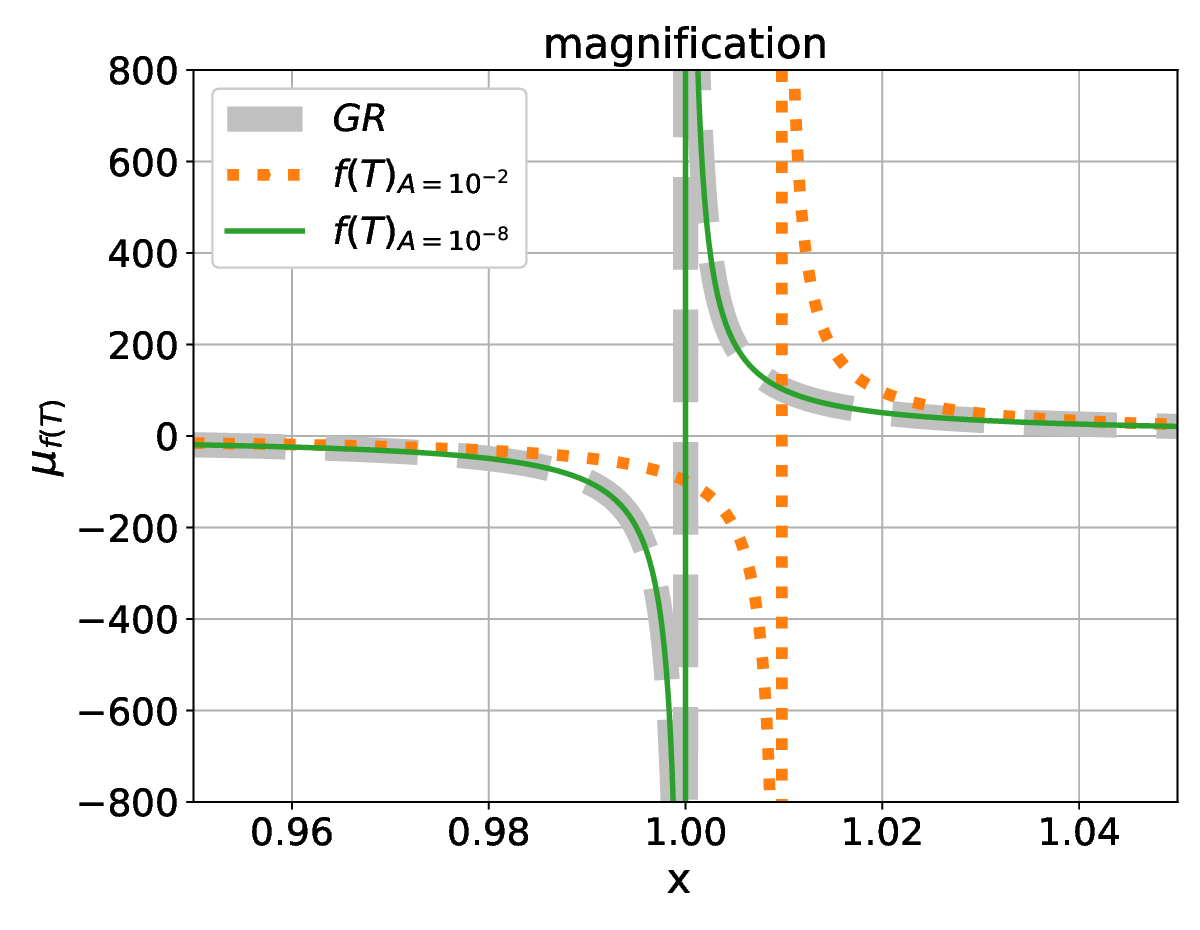}
\caption{The magnification near the critical curves on the image plane with different $A$. The grey dashed line represents that of GR. The other two coloured curves show that of $f(T)$ with different $A$. }
\label{fig:SIS_Xt_mu}
\end{figure}

Now we turn to investigate other properties of the lensing effect in $f(T)$ gravity. The weak lensing shear under $f(T)$ can be given by
%
\be
\gamma = -\left(\frac{1}{2}\frac{1}{x} +\frac{3}{2} \frac{A}{x^{3}}\right) \cdot e^{2i \varphi},
\ee
and the magnification is
\be
\mu =-\frac{x^{6}}{\left(A+x^{2}-x^{3}\right)\left(2 A+x^{3}\right)}.
\label{eq:sis-mu}
\ee
Fig.\,\ref{fig:SIS_Xt_mu} compares $\mu_{GR}$ and $\mu_{f(T)}$ with $A=10^{-2},10^{-8}$ as an example to demonstrate how the magnification shift when different values of $A$ are chosen. In GR, $\mu$ becomes infinity at exact $x=1$, i.e., the critical curve. In $f(T)$ gravity, the critical curve will shift outwards, and the larger $A$ is, the outer the critical curve. For small $A$, the green curve almost overlaps with the curve of GR. The difference between GR and $f(T)$ case becomes significant near the critical curve, where the lensing effects are strong.

\begin{figure}
  \centering
  \includegraphics[width=3.3 in]{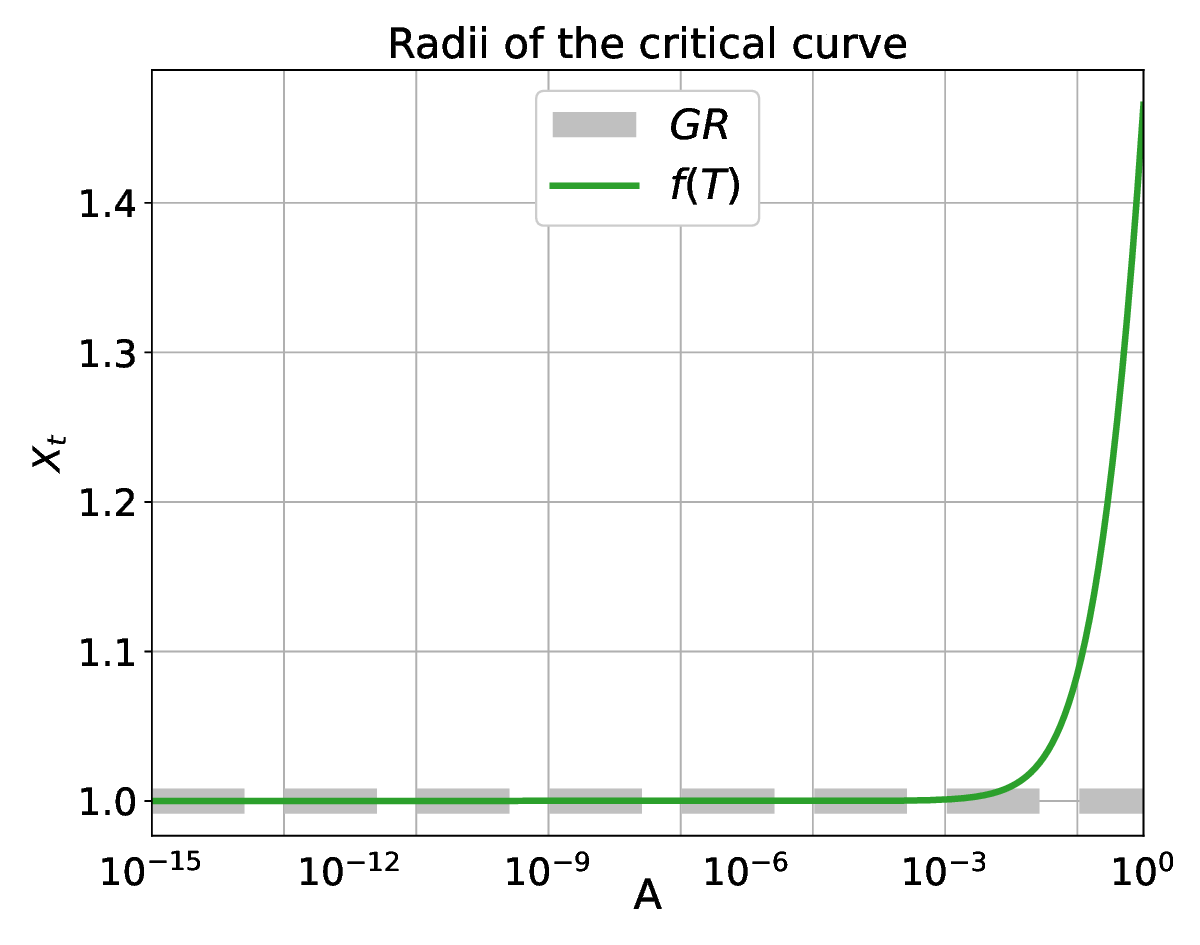}
\caption{The relation between the radii of critical curve $X_t$ and the parameter $A$ for SIS lens (Eq.\,\ref{eq:sis-xt}). The horizontal coordinate is shown in the logarithmic scale.}
\label{fig:SIS_Xt_A}
\end{figure}

The critical curve can be obtained by solving $\mu^{-1}=0$ from 
Eq.\,\ref{eq:sis-mu}. For the SIS halo is axially symmetric, the solutions give the radii of the critical curve. Since $A > 0$, under the condition that the solution is positive and real, we obtain:
\begin{flalign}
X_{t_{f(T)}}=\frac{1}{3}[1+\frac{2^{\frac{1}{3}}}{\left(2+27 A-3 \sqrt{3} \sqrt{4 A+27 A^{2}}\right)^{\frac{1}{3}}} \nonumber\\
+\frac{\left(+2+27 A-3 \sqrt{3} \sqrt{4 A+27 A^{2}}\right)^{\frac{1}{3}}}{2^{\frac{1}{3}}}].
\label{eq:sis-xt}
\end{flalign}
When $A = 0$, the solution returns back to GR, i.e., $X_{t_{GR}}=1$. We show the relation of Eq.\,\ref{eq:sis-xt} in Fig.\,\ref{fig:SIS_Xt_A}. When $A \geq 10^{-3}$, $X_{t_{f(T)}}$ differs significantly from $X_{t_{GR}}$.

\begin{figure*}  
  \center{
  \includegraphics[width=3.0 in]{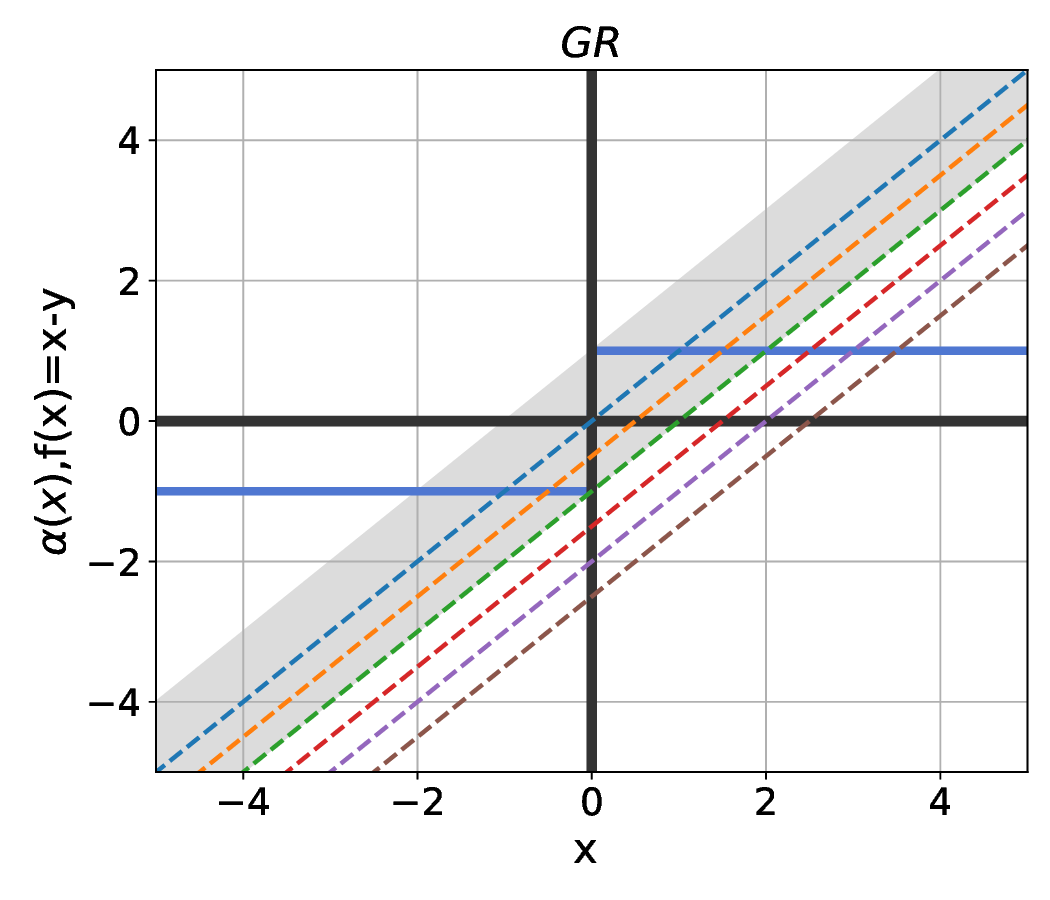}
  \includegraphics[width=3.0 in]{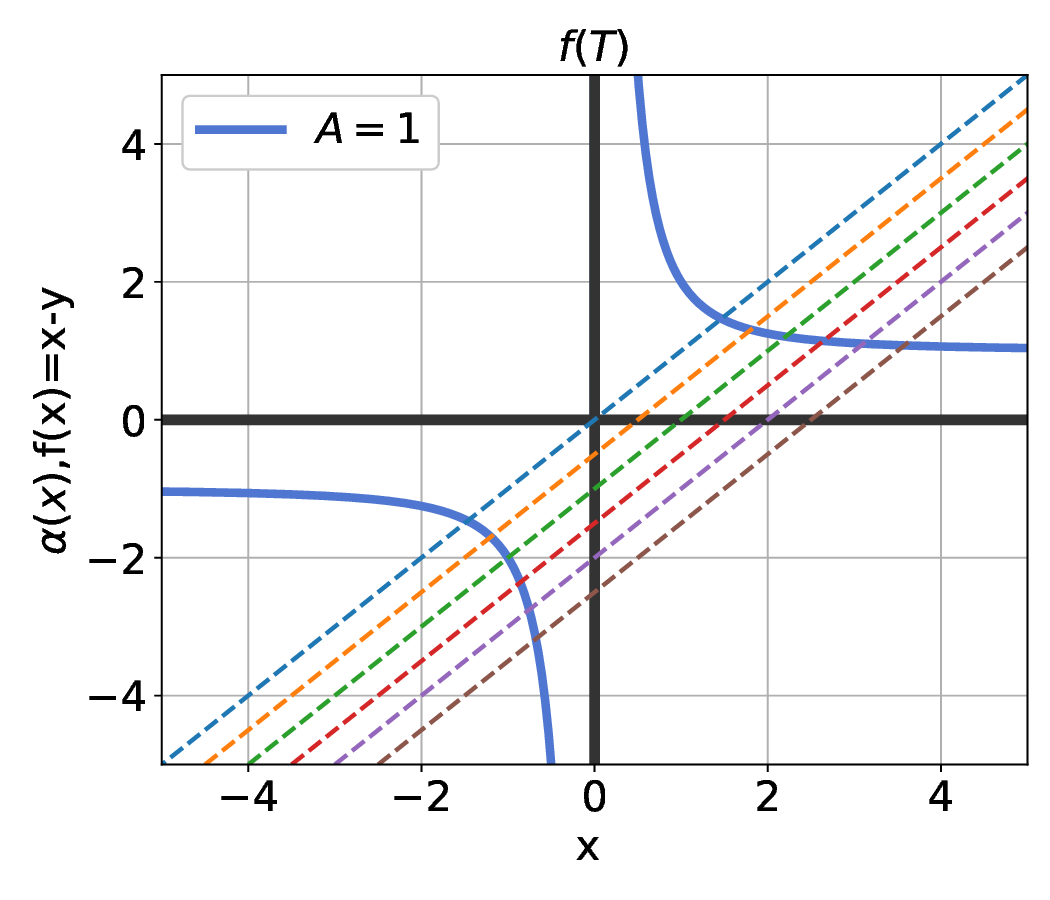}}
  \caption{Image diagram for SIS halo under GR (left panel) and $f(T)$ (right panel). The solid blue lines represent the deflection angle $\alpha(x)$. The coloured dashed lines show the function $f(x)=x-y$ for a range of $y \in [0, 2.5]$, the corresponding $y$ values from top to bottom are $y= 0,0.5,1,1.5,2,2.5$. According to lens equation Eq.\,\ref{eq:lens equation}, intersections of the dashed and solid line indicate the positions of images. While the grey area in the left panel is where $\alpha(x)$ and $f(x)$ have two intersections, i.e. lensing can generate two images in this region. }
  \label{fig:Cut}
\end{figure*}

\begin{center}
\begin{figure}
  {\includegraphics[width=3 in]{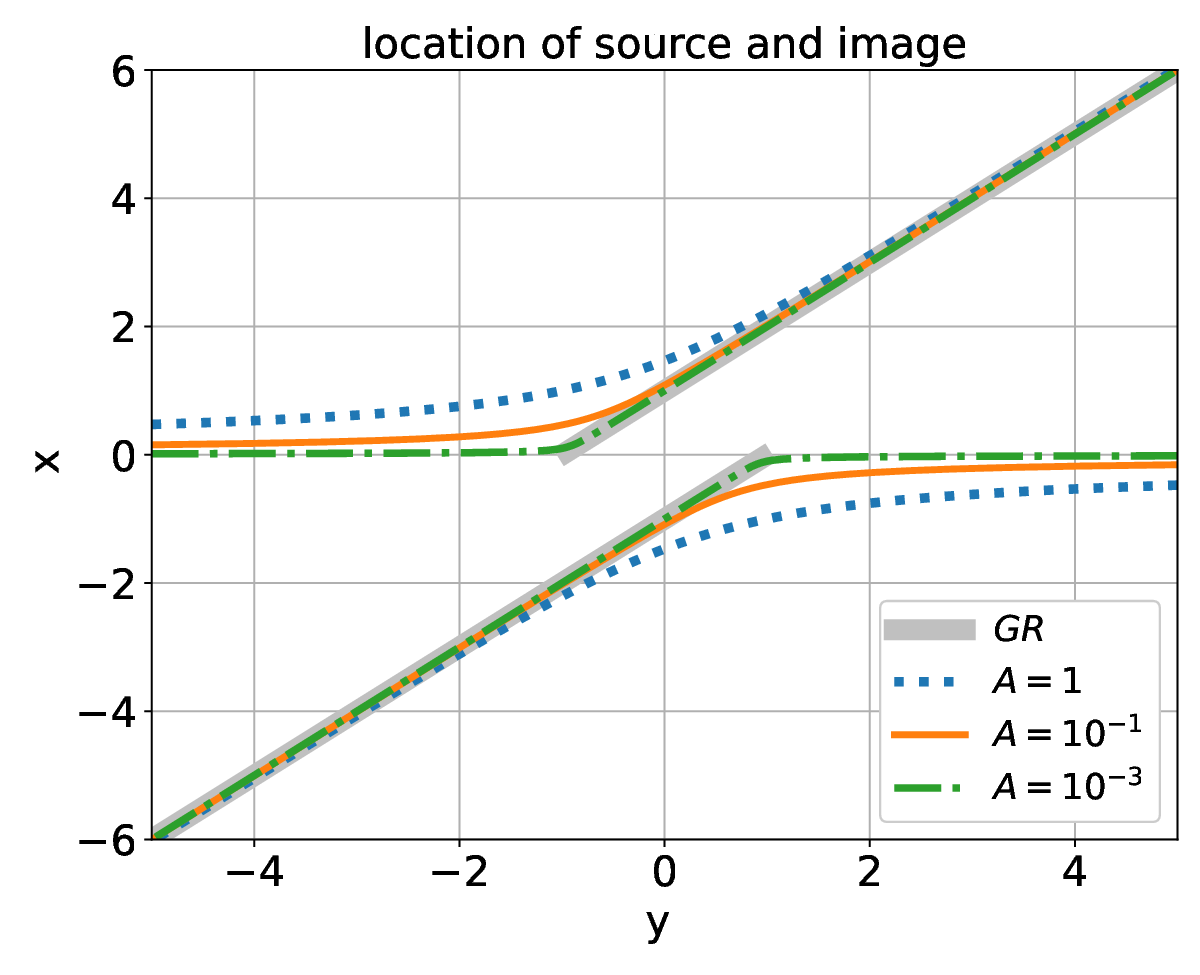}}
  \caption{Image diagram for location of source and image. The abscissa $y$ represents the position of the source, and the ordinate $x$ represents the position of the corresponding image. The $x-y$ relation under $f(T)$ gravity is marked by coloured lines, and the gray solid line is the relation under GR for reference.}
  \label{fig:SIS_xy}
\end{figure}
\end{center}

In Fig.\,\ref{fig:Cut}, we compare the image positions between the GR (left panel) and the $f(T)$ (right panel). In GR, the SIS halo can generate two images when the source is within the caustics, i.e., $|\vec{\beta}|<\theta_E$. While in $f(T)$, the lens can always generate two images in principle, even if the source is outside the caustic. However, when the secondary image is very close to the lens, it will be strongly demagnified. Furthermore, Fig.\,\ref{fig:SIS_xy} shows the image position corresponding with the different source locations in $f(T)$ and GR lensing system. For GR, there is a "jump" in the $x-y$ relationship, which is absent in the $f(T)$ case. In the two solutions of $x$, one is close to that of GR, while the other approaches $0$ as $|y|$ increases.

\subsection{SIE model}
\label{sect:SIE model under f(T) gravity}
A more general mass model for the lens galaxies is the Singular Isothermal Ellipsoid (SIE). It can be generalized by introducing elliptical coordinates $\vec{x}'=(x_1, fx_2)$, where $f$ is the axis ratio between the minor and major axis \cite[][]{Kassiola1993EllipticMD,Kormann1994IsothermalEG}. We can also write the radial coordinates of the SIE model by $x'\equiv|\vec{x}'|=x\Delta(\varphi)$, with $\Delta(\varphi)\equiv\sqrt{\cos^{2}\varphi+f^{2}\sin^{2}\varphi}$.
Then the convergence can be written by
\be
\kappa_{f(T)} =\sqrt{f}\left[\frac{1}{2 x \Delta(\varphi)}-\frac{A}{2 x^{3} \Delta^{3}(\varphi)}\right] .
\label{eq:sie-kappa}
\ee
The lensing potential can be found by solving the Poisson equation
\be
\frac{\partial^{2} \Psi_{f (T)}}{\partial x^{2}}+\frac{1}{x} \frac{\partial \Psi_{f (T)}}{\partial x}+\frac{1}{x^{2}} \frac{\partial^{2} \Psi_{f(T)}}{\partial \varphi^{2}}= 2\kappa_{f(T)}.
\ee
By solving the Poisson equation (see \hyperref[SIE lensing potential under $f(T)$ gravity]{Appendix} for more details), we get the effective SIE lensing potential for $f(T)$ gravity
\begin{flalign}
\Psi_{f(T)}
&=x \cdot \frac{\sqrt{f}}{f^{\prime}}\Bigg[\sin \varphi \arcsin \left(f^{\prime} \sin \varphi\right)\\
&+\cos \varphi \operatorname{arcsinh}\left(\frac{f^{\prime}}{f} \cos \varphi\right)\Bigg] \nonumber
-\frac{1}{x} \cdot \frac{A\sqrt{f}}{f^2} \cdot \Delta(\varphi), 
\end{flalign}
where $f^{\prime} = \sqrt{1-f^2} $.
The two components of the deflection angle are
\begin{align}
\alpha_{f(T)1}(x,\varphi)&=\frac{\sqrt{f}}{f^{\prime}} \operatorname{arcsinh}(\frac{f^{\prime}}{f} \cos \varphi) \nonumber \\
&+\frac{A \sqrt{f}}{f^{2} x^{2}} \cdot \cos\varphi [\Delta(\varphi)-\frac{\sin^2 \varphi {f^{\prime}}^{2}}{\Delta(\varphi)}], \nonumber \\ 
\mathrm{and}\quad\alpha_{f(T)2}(x,\varphi)&=\frac{\sqrt{f}}{f^{\prime}} \operatorname{arcsin}({f^{\prime}} \sin \varphi) \nonumber \\
&+\frac{A \sqrt{f}}{f^{2} x^{2}} \cdot \sin\varphi [\Delta(\varphi)+\frac{\cos^2 \varphi {f^{\prime}}^{2}}{\Delta(\varphi)}],
\label{eq:sie-alpha}
\end{align}
and the shear components are
\begin{flalign}
\gamma_{f(T)1}&=-\frac{\sqrt{f}}{\Delta}\frac{\cos 2\varphi }{2 x}+ \frac{A}{4f^{\frac{3}{2}}x^{3}\Delta^{3}}  {\left[-1+f^4-3(1+f^4)\cos 2\varphi \right.} \nonumber \\
&\left. - 3 (1 - f^4) \cos 4\varphi - (1 - f^2)^2 \cos 6\varphi \right], \nonumber \\
\gamma_{f(T)2}&=-\frac{\sqrt{f}}{\Delta}\frac{\sin 2\varphi }{2 x}+ \frac{A}{4f^{\frac{3}{2}}x^{3}\Delta^{3}}  {\left[(-4 + 2 f^2 - 4 f^4)\sin 2\varphi \right.} \nonumber \\
&\left.-3(1-f^4)\sin4\varphi -(1-f^2)^2\cos4\varphi \cdot 2 \sin2\varphi \right].
\label{eq:sie-gamma}
\end{flalign}
$\gamma_1$ and $\gamma_2$ respectively denote the components of shear along the horizontal and vertical coordinates.

%
%
%
%
%


\section{Gravitational and plasma lensing effects}
\label{sect:plasma lensing}
The difference in lensed image position between GR and $f(T)$ is small, making it difficult to distinguish the $f(T)$ theory from a strongly lensed galaxy or QSO. A feasible candidate is the lensed FRB, which has been proposed to study cosmology and fundamental physics \cite[e.g.][]{Dai2017ProbingMO,Li2017StronglyLR,Wucknitz2020CosmologyWG,Chen2021FRBsLB}. Nevertheless, the propagation of the radio signal is affected by the ionized medium, i.e., plasma, leading to non-negligible plasma lensing effects that introduce intriguing changes \cite[e.g.][]{Tsupko2014GravitationalLI,Tsupko2012OnGL,Er2022TheEO,2023arXiv231110615P}.
Thus, to achieve more precise predictions, we investigate the lensing of FRBs within the framework of $f(T)$ theory combined with plasma lensing. 

The plasma lensing effect primarily depends on the plasma density (mainly free electrons) and observational wavelength \cite[e.g.][]{BisnovatyiKogan2015GravitationalLI, Kumar2022GravitationalLI, Cordes2017LensingOF}. The plasma density in the lens galaxy is not well constrained from observation yet.
We simply adopt two electron density models on the galactic scale in this work, the power-law and the exponential ones \cite{2003ARA&A..41..191M,2010ApJ...710L..44G}. The parameters of these two plasma density models are shown in Table \ref{tab:plasma lensing}, with further details available in \cite[][]{Er2017TwoFO}. Additionally, we introduce the column-density power-law model for simplicity in our calculations. In Table \ref{tab:plasma lensing}, we summary the expressions of the parameters for three density models.
For power-law plasma lens, $N^{pl,c}_0$ and $N^{pl,v}_0$ are the electron column/volume density at a characteristic radius $r = R_0$ or $\theta = \theta_R$. And we recall that $\theta\equiv|\vec{\theta}|$ and $r\equiv(\xi^2+Z^2)^{1/2}$. For an exponential plasma lens, $N^{exp}_0$ represents the maximum electron column density within the lens and $\sigma$ is the angular size of the lens. $\theta_0$ represents the characteristic angular scale for plasma lensing, which is an analogy to the Einstein radius in gravitational lensing. $\lambda$ is the wavelength and $r_e$ is the classical electron radius. The plasma lensing potential $\Psi(\theta)$ and $\Psi(x)$ are given by the projected electron density. In this work, we use the spectral index $h > 0$ and $H>0$. $\alpha(\theta)$ is the deflection angle and the $\alpha(x)$ is the dimensionless version scaled by the Einstein radius $\theta_E$. In the following part, we mark the corresponding parameters for different plasma models with an extra superscript, e.g., $\alpha_{1}^{pl,c}$ represents the component of $\alpha(x)$ for the power-law model with column-density along the $x$-axis.

The plasma clumps at small scales can cause deflection and even dominate the plasma lensing effects. This is difficult to predict and strongly depends on the line of sight. We thus consider the simply smooth model at the current study, i.e. only a plasma lens on galactic scale, and leave the analysis of small scales in future work. 

There is coupling between GR and plasma lensing \cite{BisnovatyiKogan2010GravitationalLI}, also in the $f(T)$ gravity. Such effects are tiny and thus are not included in our calculations. We treat the overall lensing potential as the sum of individual contributions from modified gravity and plasma lensing. 

\begin{table*}
\begin{center}
\begin{tabular}{rccc}
\hline
\noalign{\smallskip}
\textbf{  } &\textbf{Column-Density Power-Law lenses} & \textbf{Volume-Density Power-Law lenses} & \textbf{Exponential Lenses} \\
\noalign{\smallskip}
\hline
\noalign{\smallskip}
{$N_{e}$} & {$N^{pl,c}_0 \frac{\theta_R^H}{\theta^H}$} & {$N^{pl,v}_0 \frac{{R_0}^h}{r^h}$} & {$N^{exp}_0 \exp \left(-\frac{\theta^h}{h \sigma^h}\right) \quad$}\\ \hline
\noalign{\smallskip}
$\theta_{0}$ & {$(\frac{\lambda^2}{2\pi} \frac{D_{LS}}{{D_S}{D_{L}}} {r_{\mathrm{e}} H N_0^{pl,c} \theta_R^H})^{\frac{1}{H+2}}$} & {$\left(\lambda^2 \frac{{D_{LS}}}{D_{S} D_{L}^h} \frac{r_{\mathrm{e}} N_0^{pl,v} R_0^h}{\sqrt{\pi}} \frac{\Gamma\left(\frac{h}{2}+\frac{1}{2}\right)}{\Gamma\left(\frac{h}{2}\right)}\right)^{\frac{1}{h+1}}$} & {$\lambda\left(\frac{D_{LS}}{D_{S} D_{L}} \frac{1}{2 \pi} r_{\mathrm{e}} N_0^{exp}\right)^{\frac{1}{2}}$} \\
\noalign{\smallskip}
\hline
\noalign{\smallskip}
{$\Psi(\theta)$} & {$\frac{1}{H}\frac{\theta_0^{H+2}}{\theta^H}$} & {$\begin{cases}\frac{\theta_0^{h+1}}{(h-1)} \frac{1}{\theta^{h-1}}, & h \neq 1 \\ -\theta_0^2 \ln \theta, & h=1\end{cases}$} & {$\theta_0^2 e^ {-\frac{\theta^h}{h \sigma^h}}$}  \\
\noalign{\smallskip}
\hline
\noalign{\smallskip}
{$\Psi(x)$} & {$\frac{1}{H x^{H}} (\frac{\theta_0}{\theta_E})^{H+2} $} & {$\begin{cases}  (\frac{\theta_0}{\theta_E})^{h+1} \frac{1}{(h-1)x^{h-1}}, & h \neq 1 \\ -(\frac{\theta_0}{\theta_E})^2 \ln (\theta_E \cdot x), & h=1\end{cases}$} & {$ (\frac{\theta_0}{\theta_E})^2 e ^{-\frac{\theta_E^h}{h \sigma^h} \cdot x^h} $} \\
\noalign{\smallskip}
\hline
\noalign{\smallskip}
$\alpha(\theta) $ & {$ -\frac{\theta_0^{H+2}}{\theta^{H+1}}$} & {$-\frac{\theta_0^{h+1}}{\theta^h}$}  & {$-\theta_0^2 \frac{\theta^{h-1}}{\sigma^h} e^{-\frac{\theta^h}{h \sigma^h}}$} \\ 
\noalign{\smallskip}
\hline
\noalign{\smallskip}
$\alpha(x) $ & {$-(\frac{\theta_0}{\theta_E})^{H+2} \frac{1}{x^{H+1}}$} &  {$-(\frac{\theta_0}{\theta_E})^{h+1} \cdot \frac{1}{x^h}$}  & {$ -x^{h-1} \frac{{\theta_E}^{h-2}}{\sigma^h {\theta_0}^{-2}} e^{-\frac{x^h}{h} \cdot \frac{\theta_E^h}{\sigma^h}}$}\\
\noalign{\smallskip}
\hline
\end{tabular}
\end{center}
\caption{The expressions of parameters in three plasma lensing models: the column-density power-law model, volume-density power-law model, and exponential model from left to right respectively.}
\label{tab:plasma lensing}
\end{table*}

\subsection{\texorpdfstring{\bm{$f(T)\;$ \&}} . plasma lensing for SIS profile}
\label{sect:plasma lensing for SIS profile SIS}

\subsubsection{power-law plasma lenses}
For the SIS model, we take the power-law density for plasma with index $h=2$. Such a choice can give the same density profile as the dark matter halo. We introduce the parameter $b_{pl} = (\frac{\theta_0}{\theta_E})^{h+1}$ to indicate the strength of plasma lensing. Then the deflection angle can be written as
\be
\begin{aligned}
\alpha =\alpha_{f(T)}+\alpha^{pl,v} =1+\frac{A}{x^{2}}-\frac{b_{pl}}{x^2}.
\label{eq:SIS_alpha_fTpl}
\end{aligned}
\ee
From this expression, it is evident that the modification due to $f(T)$ gravity and plasma lensing can be distinguished into the following three cases: when $A > b_{pl}$, the modification is dominated by $f(T)$ gravity; when $A = b_{pl}$, the effects of $f(T)$ and plasma cancel out; and when $A < b_{pl}$, the modification is dominated by the plasma lensing effect.

The time delay is
\be
\begin{aligned}
td &=D_t \theta_E^2\left[\frac{1}{2}\left(\vec{x}-\vec{y}\right)^2-\Psi_{f(T)}(x)+\frac{1}{(1+z_L)^2} \Psi^{pl,v} (x)\right] \\
&=D_t \theta_E^2\left[\frac{1}{2}(\vec{x}-\vec{y})^2-\left(x-\frac{A}{x}\right)+\frac{1}{(1+z_L)^2} \frac{b_{pl}}{x}\right],
\end{aligned}
\ee
It consists of the following three components: path deflection, $f(T)$ gravity, and plasma environment, respectively. $D_t$ is defined by $D_t\equiv D_L D_s\left(1+z_L\right) / (cD_{L S})$. The total magnification $\mu$ is a combination of plasma and gravitational lensing effects, which is given by
\be
\begin{aligned}
\mu=&\left[\left(1 -\frac{\partial(\alpha_{f(T)}+\alpha^{pl,v})}{\partial x}\right)\left(1-\frac{\alpha_{f(T)}+\alpha^{pl,v})}{x}\right)\right]^{-1}\\
=&\left[\frac{{\left(-A + b_{pl} - x^2 + x^3\right) \left(2A - 2b_{pl} + x^3 \right)}}{{x^6}} \right]^{-1}.
\end{aligned}
\ee
\subsubsection{Exponential plasma lenses}
\label{sect:exponential plasma lenses SIS}

\begin{figure}
  {\includegraphics[width=3.1 in]{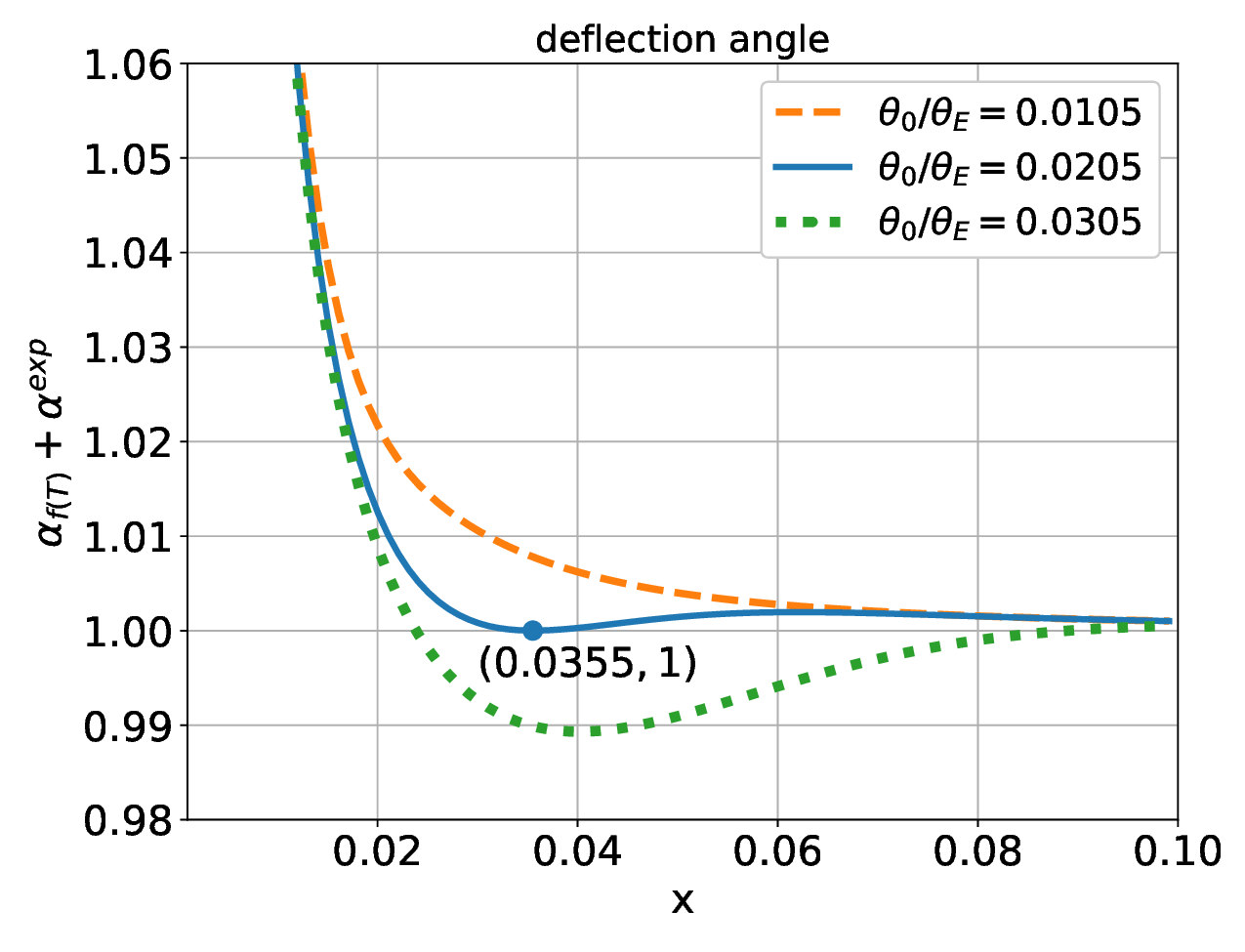}}
  \caption{Deflection angle $\alpha^{exp}+\alpha_{f(T)}$ of $x$ with different $\theta_0$. The blue dot marked in the figure is the minimum value of $\alpha^{exp}+\alpha_{f(T)}$ when $\theta_0/ \theta_E = 0.0205$.}
  \label{fig:alpha_fTexp}
\end{figure}

The derivation of exponential plasma with $f(T)$ gravity lensing effects is similar to that of the power-law plasma model. We start from the deflection angle:
\be
\begin{aligned}
\alpha = \alpha_{f(T)} + \alpha^{exp} = 1 + \frac{A}{x^2} - x^{h-1} \frac{\theta_E^{h-2} \theta_0^{2}}{\sigma^h} \exp\left[-\frac{x^h}{h} \cdot \frac{\theta_E^h}{\sigma^h}\right].
\end{aligned}
\ee
We adopt the Gaussian model in this work, i.e., $h=2$, and define $b_{exp}\equiv \theta_0^2/\sigma^2$. 
We take the derivative to compare the modification by $f(T)$ and plasma
\be
\begin{aligned}
\frac{\partial \alpha}{\partial x}=-\frac{2 A}{x^3}+(-b_{exp} + x^2 b_{exp}^2 \frac{\theta_E^2}{\theta_0^2}) \, \exp\left(-\frac{ x^2 \theta_E^2 b_{exp}}{2\theta_0^2}\right).
\end{aligned}
\ee
Unlike the deflection angle caused by $f(T)$ gravity and power-law plasma distribution, which can be simply evaluated by given $A$ and $b_{pl}$, the deflection angle here shows complicated behaviour.
To simplify the calculation, we take $\sigma = \theta_0$ ($b_{exp}=1$) as an example. 
In Fig.\,\ref{fig:alpha_fTexp}, we show the deflection angle, where we take $A = 10^{-5}$, $z_{S}=0.08$, $z_{L}=0.01$, $\sigma_{v} = 200\ \mathrm{km/s}$, the corresponding $\theta_E \approx 1$ arcsec. For plasma, we use $N_{0}^{exp}= 10^3$ pc\,cm$^{-3}$. The curves with different colour present the deflection angles at different frequencies, which are $195$MHz, $290$MHz, and $565$MHz. The corresponding $\theta_0/\theta_E$ are $0.0305, 0.0205$ and $0.0105$ respectively. When $\theta_0/\theta_E<0.0205$, the deflection angle is always greater than $1$; when $\theta_0/\theta_E=0.0205$, the deflection angle has a minimum value (blue dot) at which the deflection caused by the plasma cancels out the extra deflection by $f(T)$. Then as $x$ increases, the deflection angle exhibits a small peak and then gradually decreases. When $\theta_0/\theta_E>0.0205$ ($\nu<290$MHz), the plasma effect becomes more significant, and the deflection angle can be smaller than that of GR. 

The time delay consists of three components
\be
\begin{aligned}
td =D_t \theta_E^2&\left[\frac{1}{2}(\vec{x}-\vec{y})^2-\Psi_{f(T)}(x)+\frac{1}{(1+z_L)^2} \Psi^{exp}(x)\right]\\
=D_t \theta_E^2&{\left[\frac{1}{2}(\vec{x}-\vec{y})^2-\left(x-\frac{A}{x}\right) \right.}  \\
&+\left. \frac{1}{(1+z_L)^2} b_{exp}\frac{\theta_0^2}{\theta_E^2}\exp\left(-b_{exp}\frac{x^2 \theta_E^2}{2 \theta_0^2}\right)\right]
\end{aligned}
\ee
The total magnification $\mu$ is composed of plasma and gravitational lensing effects:
\be
\begin{aligned}
\mu=&\left[\left(1 -\frac{\partial(\alpha_{f(T)}+\alpha^{exp})}{\partial x}\right)\left(1-\frac{(\alpha_{f(T)}+\alpha^{exp})}{x}\right)\right]^{-1}\\
=&{\left[\left(1 + b_{exp} e^{-\frac{b_{exp} x^2 \theta_E^2}{2 \theta_0^2}} - \frac{A}{x^3} - \frac{1}{x}\right) \right.}  \\
\cdot & \left. \left(1 + \frac{2A}{x^3} + b_{exp} e^{-\frac{b_{exp} x^2 \theta_E^2}{2 \theta_0^2}}(1 - b_{exp} \frac{x^2 \theta_E^2}{\theta_0^2}) \right) \right]^{-1}.
\end{aligned}
\ee

\subsubsection{Results}
\begin{figure*}
\centering
\includegraphics[width=0.92\textwidth]{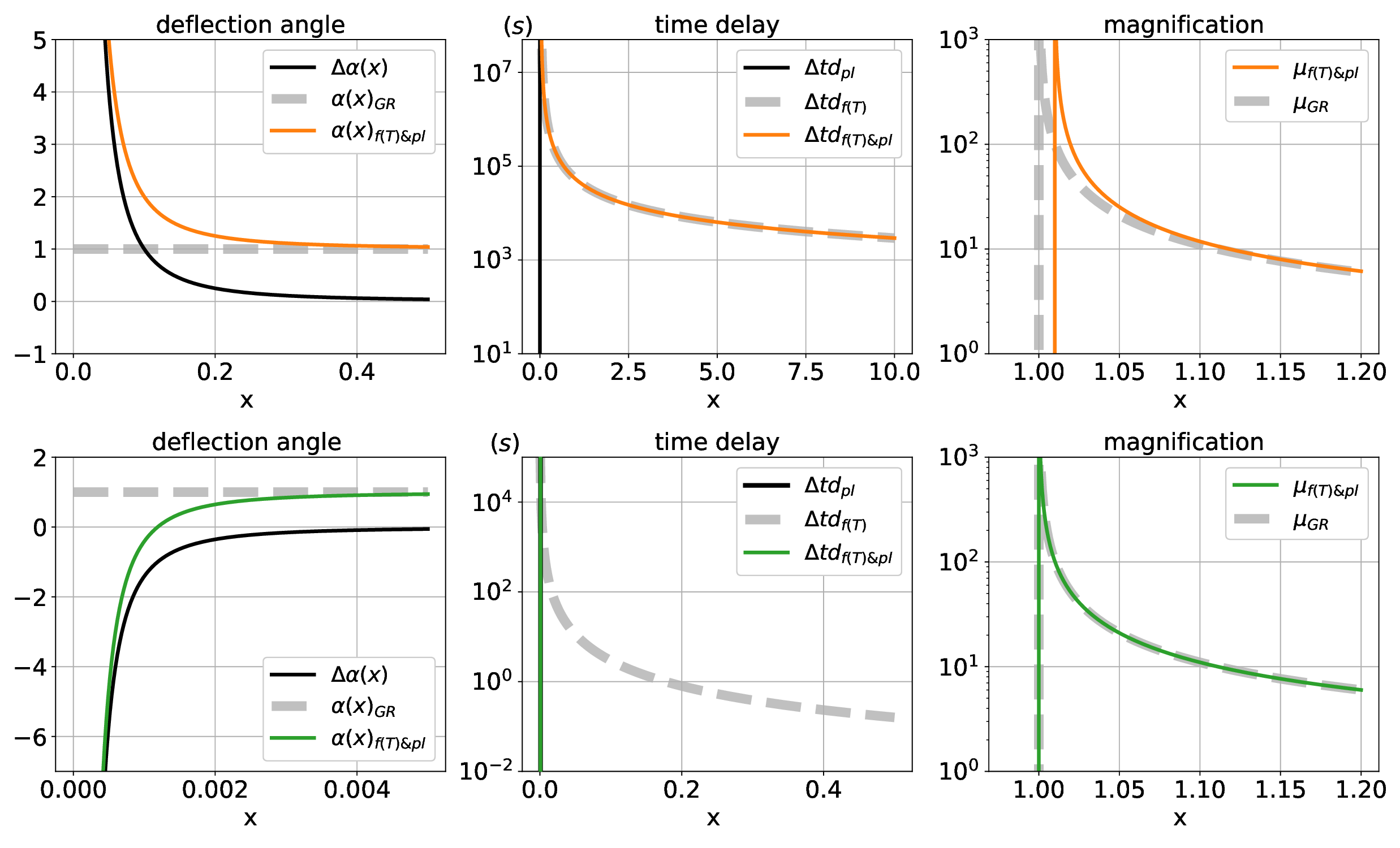}  
\caption{Comparison of lensing effects between $f(T)$ gravity, w/o plasma lensing. SIS halo is used for the mass of the lens, and the power-law model is used for the plasma. The top (bottom) row depicts the quantifies for the $f(T)$ lensing with parameter $A=10^{-2}$ ($A=10^{-8}$). $\Delta \alpha(x)$ refers to $\alpha(x)_{f(T) \& pl}-{\alpha(x)}_{GR}$, and $\Delta td_{f(T) \& pl}$, $\Delta td_{pl}$, $\Delta td_{f(T)}$ respectively represents the value of $td_{f(T) \& pl}-td_{GR}$, $td_{pl}-td_{GR}$, $td_{f(T)}-td_{GR}$. The horizontal scale of each panel is different for better visibility.} 
 \label{fig:SIS_pl}
\end{figure*}

\begin{figure*}
\includegraphics[width=0.90\textwidth]{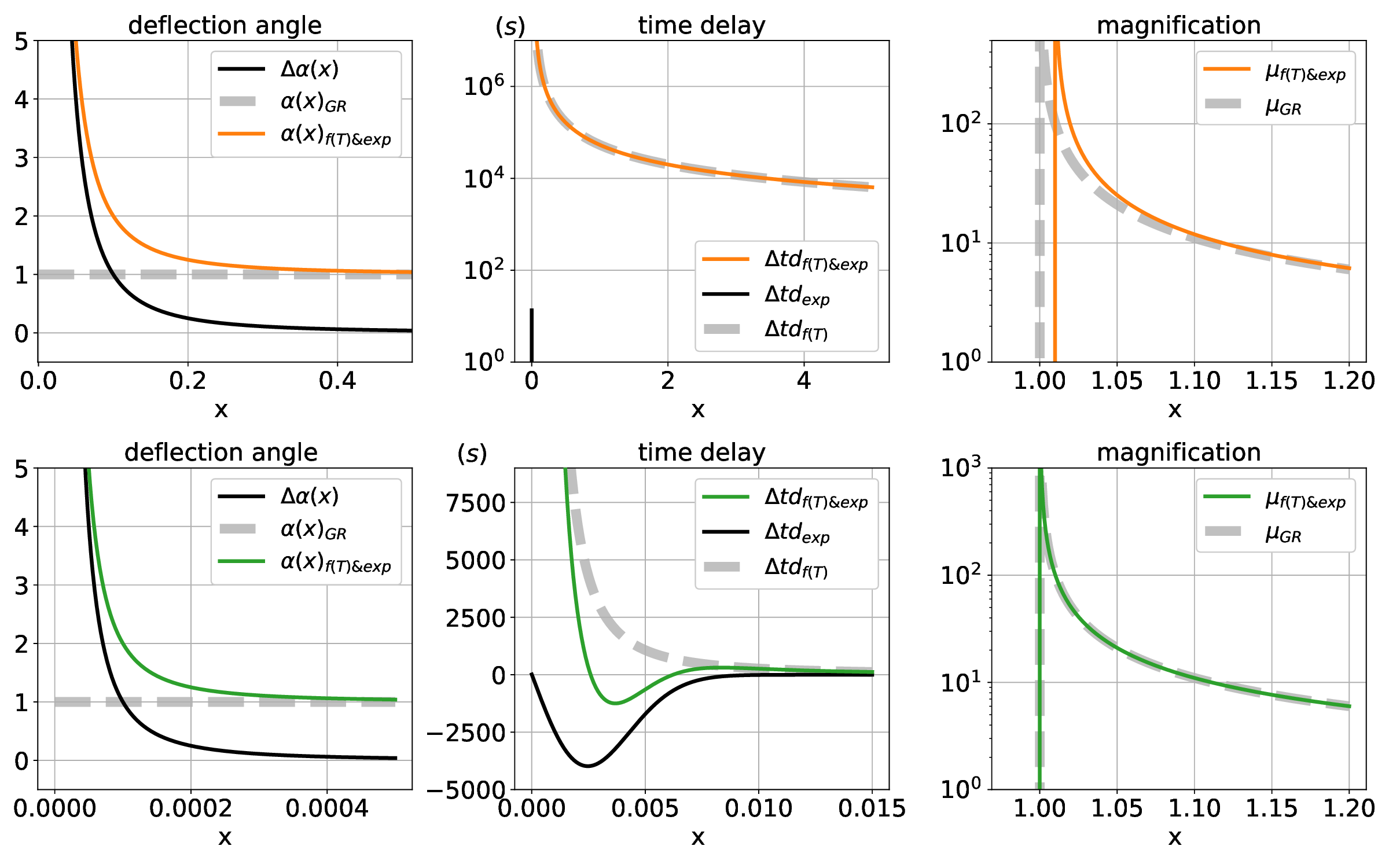}  
\caption{Same as Fig.\,\ref{fig:SIS_pl} but for Gaussian plasma density profile. In top (bottom) panel, $A=10^{-2}$ ($A=10^{-8}$) is used for $f(T)$.} 
\label{fig:SIS_exp}
\end{figure*}

We present the lensing effect of an SIS halo with plasma, including the defection angle, time delay, and magnification in Fig.\,\ref{fig:SIS_pl} and Fig.\,\ref{fig:SIS_exp}. We adopt the velocity dispersion $\sigma_v=250\ \mathrm{km/s}$ for the galaxy halo. The redshifts of the lens and the source are $z_L = 0.1$ and $z_S = 0.6$, respectively. 
The parameters of the plasma lens are shown in Table.\,\ref{tab:model parameter}. We arbitrary choose a high plasma density to exaggerate the plasma lensing effects.

In Fig.\,\ref{fig:SIS_pl}, the lensing properties for power-law plasma model are depicted. One can easily understand the difference between the top and bottom panels: the positive $\Delta\alpha(x)$ is dominated by the $f(T)$ gravity, while the negative one is by plasma. Moreover, the difference becomes greater with smaller $x$. In the middle panels, we can see the different extra delays induced by the $f(T)$ gravity and plasma. The $f(T)$ gravity leads to a positive extra delay, while the plasma causes a negative extra delay (which cannot be shown in the figure, black and green solid lines in the bottom middle panels). The reason for that is the plasma reduce the deflection angle, and then the geometric delay. When $A$ is larger than $b_{pl}$, $f(T)$ dominates the change in time delay (orange solid line in the top middle panel). From the right panels, the size of the critical curves is enlarged mainly by $f(T)$ gravity. The effect of plasma lensing is not significant in the magnification.

\begin{table}
  \renewcommand\arraystretch{2}
  \center\begin{tabular}{c|c|c|c}
    \hline
    {$f_{obs}$} & {$N_0^{pl,v}$} & {$R_0^{pl,v}$} & {$N_0^{exp}$} \\ 
    \hline
    \hline
    \rule{0pt}{3ex}
    $375\;$ MHz & $10^{-3}\,\mathrm{cm^{-3}}$ & 10 \,kpc & $500\,\mathrm{pc\,cm^{-3}}$\\
    \hline
  \end{tabular}
  \caption{The parameters for the plasma density model, the observational frequency $f_{obs}=c/\lambda$. $N_0^{pl,v}$ is the volume density of power-law model at $R_0^{pl,v}$. }
  \label{tab:model parameter}
\end{table}

In Fig.\,\ref{fig:SIS_exp}, we change to the exponential plasma model. In both cases (large and small $A$), the deflection angle is dominated by the $f(T)$ gravity and remains positive, since $\theta_0/\theta_E \ll 1$. The extra time delay by plasma lensing is still negative, but smaller than that of power-law. In the bottom middle panel, we are able to show $\Delta td_{exp}$ (black solid line), which show a minimum delay near $x\sim\sigma$. The total time delay $\Delta td_{f(T)\& exp}$ (green solid line) is dominated by plasma near $x\sim \sigma$, and by $f(T)$ at the other $x$.
The magnification of the exponential models is similar to that with the power-law profile: the difference is mainly caused by $f(T)$ and becomes significant near $x=1$.
However, the difference in magnification is small and difficult to distinguish in real observation.

\subsection{\texorpdfstring{\bm{$f(T)\;$ \&}} .  plasma lensing for SIE profile}
\label{sect:plasma lensing for SIE profile SIE}

In a general case, e.g., the SIE halo, it is reasonable to assume that the free electron shares a similar elliptical distribution to the dark matter halo. Therefore, we require the lensing properties of plasma to be similar with that of SIE dark matter halo. Again we generalized the coordinate by $x'\equiv|\vec{x}'|=x\Delta(\varphi)$. One can find the detailed analysis of elliptical plasma lensing in \cite[][]{Er2019TwoFO}.

\subsubsection{Power-law plasma lenses}
\label{sect:power-law plasma lenses}

We use the capitalized $H$ for the two-dimensional power index to distinguish that in three dimensions. The free electron density $N_{e}$ is given by
\be
\begin{aligned}
N_e(x,\varphi)=N_0^{pl,c} \frac{\theta_R^H}{\theta_E^H} \frac{1}{x^H \Delta^H(\varphi) }.
\end{aligned}
\ee
Thus, the plasma lensing potential is
\be
\begin{aligned}
{\Psi}^{pl,c}(x,\varphi)=\frac{1}{H} \cdot \frac{\theta_0^{H+2}}{\theta_E^{H+2}} \cdot \frac{1}{x^H \Delta^H(\varphi) }.
\end{aligned}
\ee
Here we define $B_{pl}\equiv\frac{\theta_0^{H+2}}{\theta_E^{H+2}}$, the corresponding plasma lensing deflection angles are
\be
\begin{aligned} 
& \alpha_1^{pl,c}(x,\varphi)=-\frac{B_{pl}}{x^{1+H}\Delta^H(\varphi)} \cdot \cos \varphi,\\
& \alpha_2^{pl,c}(x,\varphi)=-\frac{B_{pl}}{x^{1+H}\Delta^H(\varphi)} \cdot \sin \varphi.
\end{aligned}
\ee
After combining with Eq.\,\ref{eq:sie-alpha}, the total dimensionless deflection angle can be expressed as
\be
\alpha_j(x,\varphi)=\alpha_{f(T)j}(x,\varphi) + \alpha_j^{pl,c}(x,\varphi), \quad {\rm with}
\quad j=1,2.
\ee
The shear and convergence for power-law plasma lensing can be given by
\be
\begin{aligned}
& \gamma^{pl,c} (x,\varphi)=\frac{B_{pl}}{2}\left(2+H\right)  \frac{1}{x^{H+2} \Delta^H(\varphi)} \cdot e^{2i \varphi},\\
& \kappa^{pl,c} (x,\varphi)=\frac{B_{pl}}{2}\left(1+H\right)  \frac{1}{x^{H+2} \Delta^H(\varphi)}.
\end{aligned}
\ee
%
\subsubsection{Exponential plasma lenses}
\label{sect:exponential plasma lenses SIE}

Similarly, for SIE model, the $N_{e}(x)$ can be generalized by elliptical coordinate $x$
\be
\begin{aligned}
N_e(x,\varphi)=N_0^{exp} e^{-\dfrac{\theta_E^h}{h \sigma^h} x^h \Delta^h(\varphi)},
\end{aligned}
\ee
Thus the plasma lensing potential is
\be
\begin{aligned}
{\Psi}^{exp}(x,\varphi)=\frac{\theta_0^2}{\theta_E^2} e^{-\frac{\theta_E^h}{h \sigma^h}x^h \Delta^h(\varphi)}.
\end{aligned}
\ee
Here we define $B_{exp} =\frac{\theta_0^h}{\sigma^h}$, for the Gaussian model in this work, i.e., $h=2$, the deflection angle reads
\be
\begin{aligned}
& \alpha^{exp}_1 (x,\varphi)=C^{exp}  \cdot \cos \varphi,\\
& \alpha^{exp}_2 (x,\varphi)=C^{exp}  \cdot f^2 \sin \varphi,
\end{aligned}
\ee
with
\be
\begin{aligned}
&C^{exp}\equiv- B_{exp} x e^{-\frac{1}{2} [(\frac{\theta_E}{\theta_0})^{2} B_{exp} x^2 \Delta^{2}(\varphi)]}.\\
\end{aligned}
\ee
After combining with Eq.\,\ref{eq:sie-alpha}, the total dimensionless deflection angle can be expressed as
\be
\alpha_j(x,\varphi)= \alpha_{f(T)j}(x,\varphi) + \alpha^{exp}_j(x,\varphi), \quad {\rm with}
\quad j=1,2.
\ee
And then the shear and convergence for exponential plasma lensing can be given by
\be
\begin{aligned}
\gamma^{exp}_{1}=& D^{exp} \left[\left(E^{exp}- 1\right)\left(\cos^2(\varphi) - f^4 \sin^2(\varphi)\right) + f^2 \cos(2\varphi) \right],
\\
\gamma^{exp}_{2}=& D^{exp} E^{exp} f^2 \sin(2\phi),
\end{aligned}
\ee
and
\be
\begin{aligned}
\kappa^{exp}&= D^{exp} \left[E^{exp} - 1 \right] \Delta^4,
\end{aligned}
\ee
with
\be
\begin{aligned}
&D^{exp}\equiv \frac{1}{2 \Delta^2} B_{exp} e^{-\frac{1}{2}{B_{exp} x^2 (\frac{\theta_E}{\theta_0})^2 \Delta^2}},\\
&E^{exp} = B_{exp} x^2 \left(\frac{\theta_E}{\theta_0}\right)^2 \Delta^2.\\
\end{aligned}
\ee
respectively. 
\subsection{A mock lensed FRB with SIE lens}
\label{sect:mock lensd FRB(SIE)}

\begin{center}
\begin{figure*}  
  {\mbox{\includegraphics[width=3.5 in]{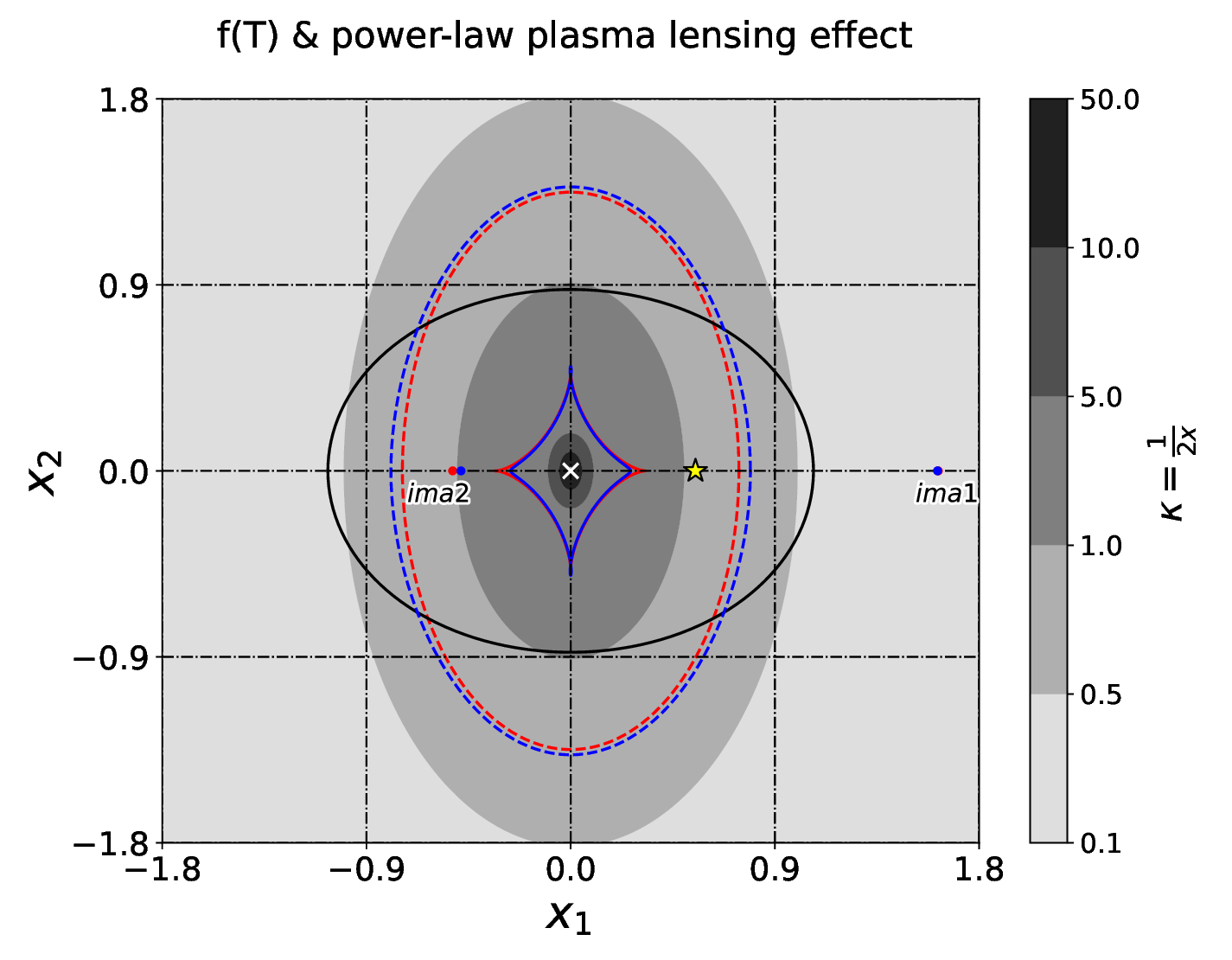}}{\includegraphics[width=3.5 in]{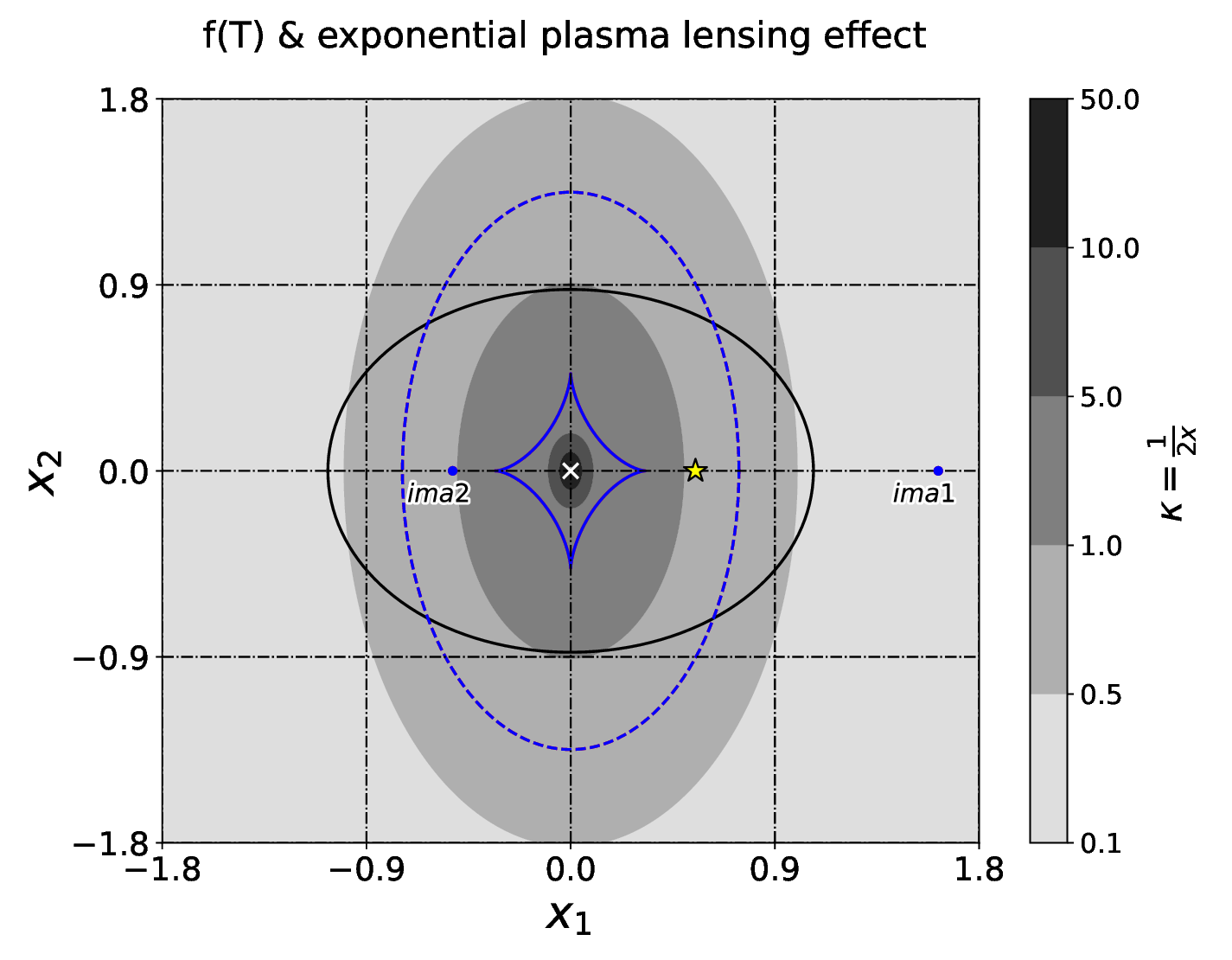}}\\{\includegraphics[width=3.5 in]{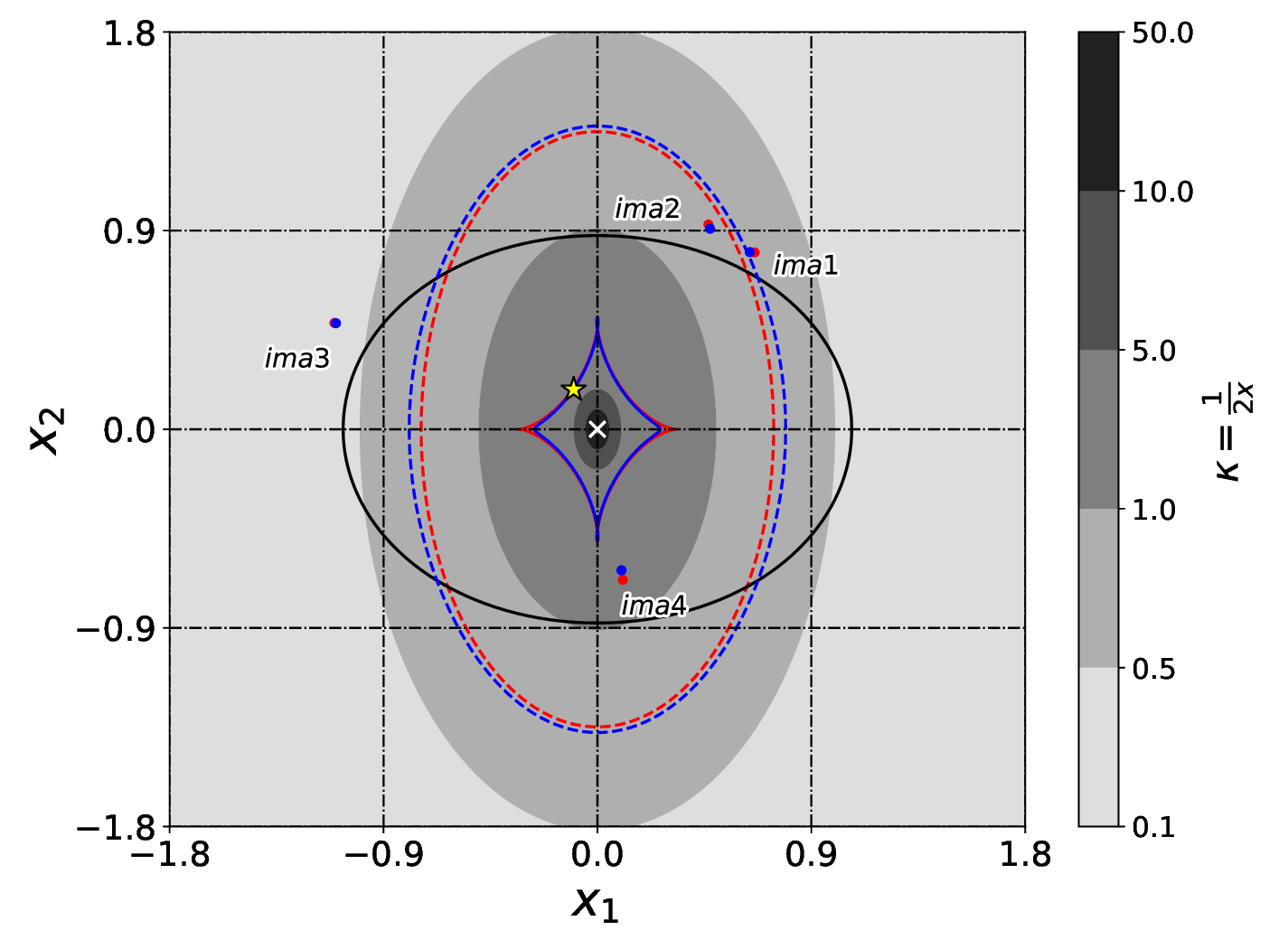}}{\includegraphics[width=3.5 in]{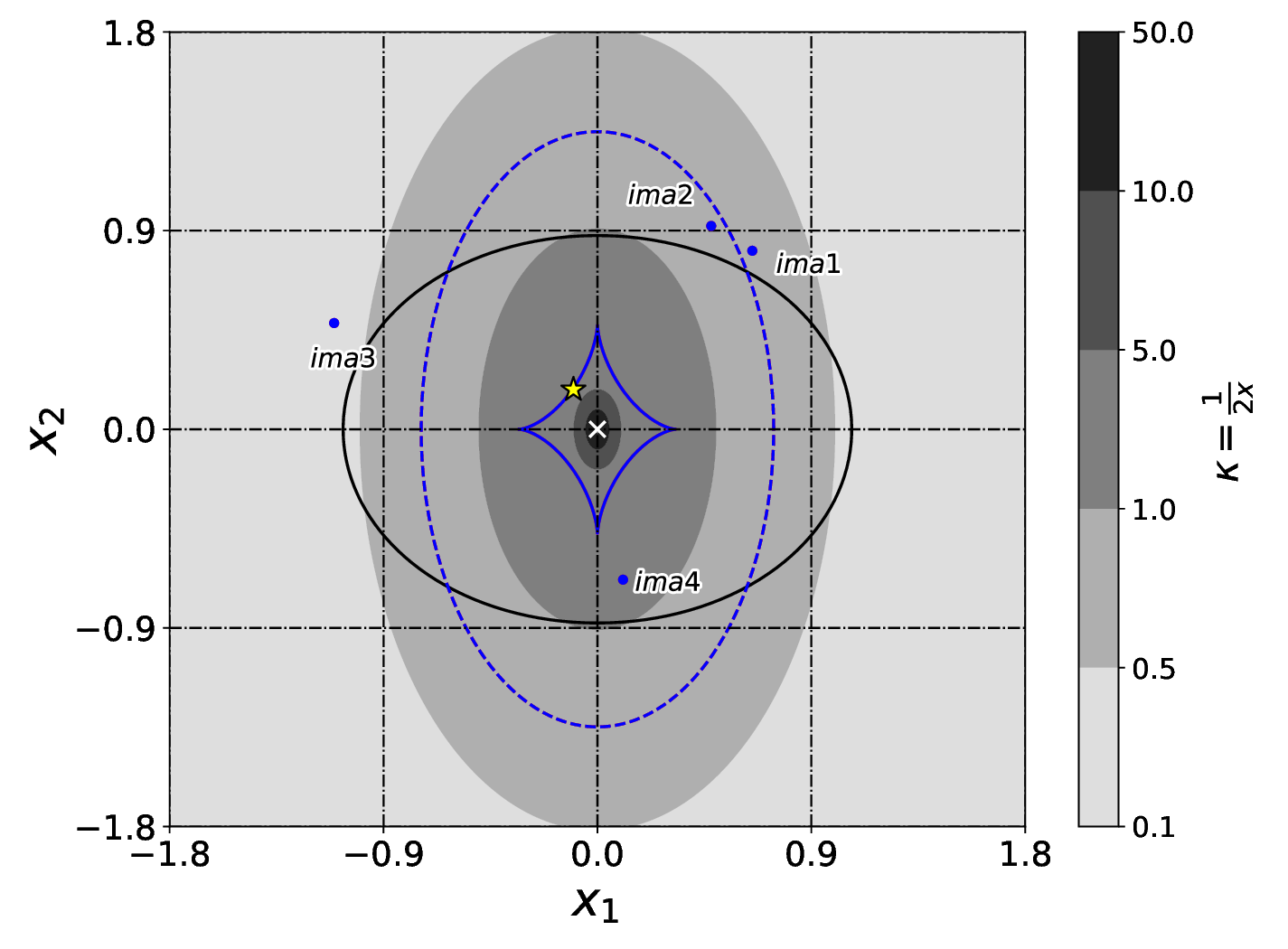}}\\{\includegraphics[width=3.5 in]{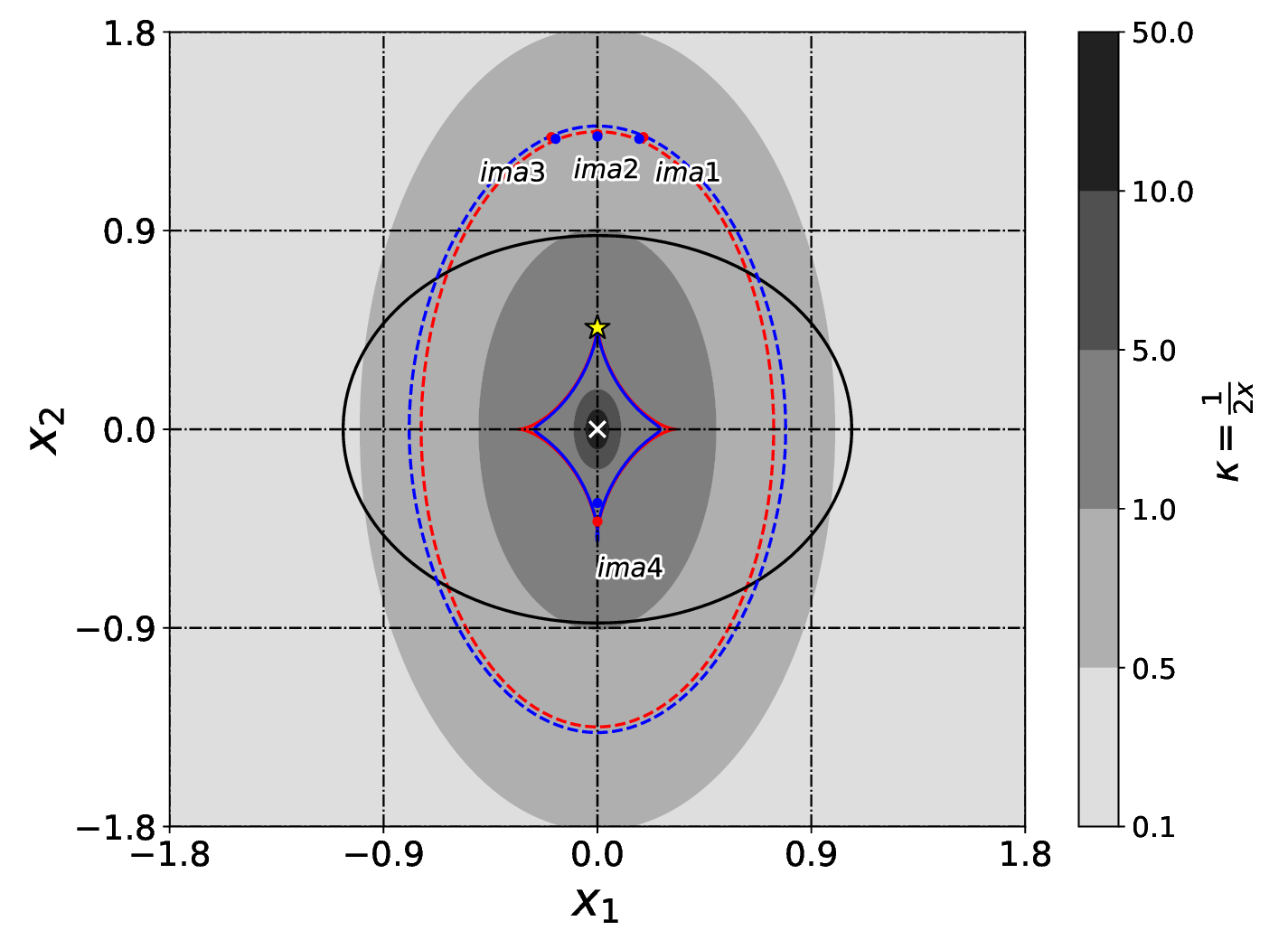}}{\includegraphics[width=3.5 in]{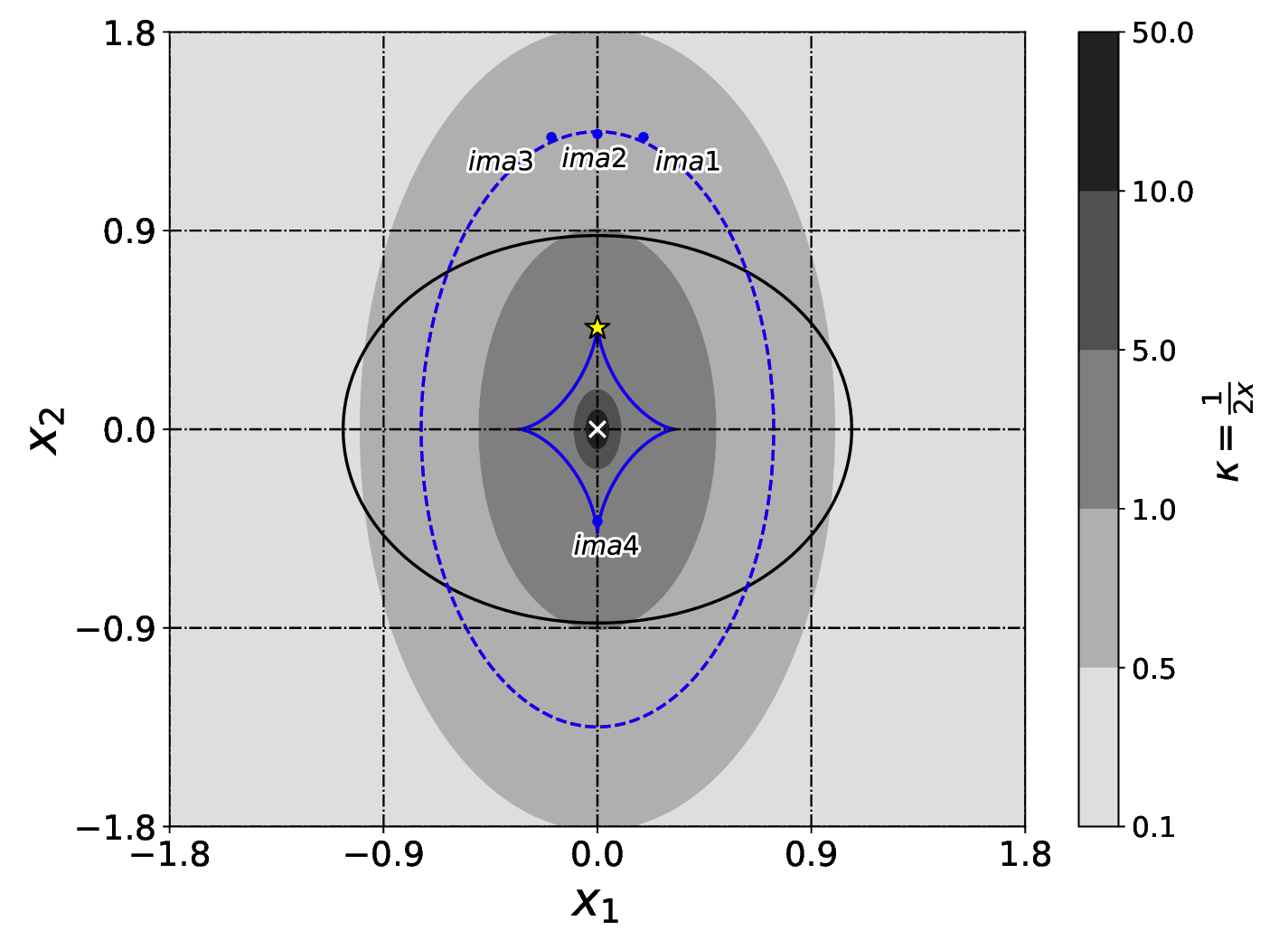}}}
  \caption{The mocked lensed FRB: in the left (right) panels, the power-law plasma model (Gaussian plasma model) is used. The grayscale indicates the surface mass density $\kappa$. The lens is located at the origin (white cross). For both two plasma models, locations of the source  (yellow star) are $(+0.55,+0.0),(-0.10,+0.18),(0,+0.46)$ from top to bottom, corresponding to the source located on the horizontal coordinate, near the fold and inside the cusp respectively. The red and blue dots mark the image positions under GR and $f(T)$ gravity respectively. The dashed line, the solid line in red (blue) marks the critical curve and caustic under GR gravity ($f(T)$ gravity) with plasma. The solid black line represents the cut for the GR case (as stated above, there is no cut for $f(T)$ gravity lensing effect).}
\label{fig:SIE mocked FRB}
\end{figure*}
\end{center}

We adopt a toy lensing model to simulate strongly lensed FRBs and compare them in GR and $f(T)$ gravity. In this section, we take $H=1$ for power law plasma density and $h=2$ for the exponential plasma model (Gaussian model). For the power-law plasma lensing, the lensing effects of plasma with column density $H=h-1$ are equivalent to the volume density with power index $h$ \cite[][]{Er2019TwoFO}. For the gravitational lens, we take $z_{L} = 0.005, z_{S} = 0.012, \sigma_v = 250 \mathrm{km/s}$, and the ellipticity is taken to be 0.55, which is consistent with the plasma lens. Other parameters related to the plasma lens are shown in Table.\,\ref{tab:SIE parameter}. Different source FRB locations are used for comparison. In Fig.\,\ref{fig:SIE mocked FRB}, we place the source (yellow stars) on the $\text{x}_1$-axis, near the fold, and close to the cusp from top to bottom. There are small differences in lensed images between the GR and $f(T)$ gravity (shown by the red and blue points).

\begin{table}
  \renewcommand\arraystretch{2}
  \center\begin{tabular}{c|c|c|c|c}
    \hline
    {$f$ (axis ratio)} & {$f_{obs}$}  & {$R_0^{pl,c}$} & {$N_0^{pl,c}$} & {$N_0^{exp}$} \\ 
    \hline
    \hline
    \rule{0pt}{3ex}
    0.55 & $375\,$MHz & $8\,$kpc & $500\,\mathrm{pc\,cm^{-3}}$ & $2000\,\mathrm{pc\,cm^{-3}}$\\
    \hline
  \end{tabular}
  \caption{The parameters of plasma density in the SIE lens model.}
  \label{tab:SIE parameter}
\end{table} 
In the left panels of Fig.\,\ref{fig:SIE mocked FRB}, we have $A=0.001$ for $f(T)$ gravity and $B_{pl}=0.011$ for plasma lensing. The critical curve with plasma (blue dashed) is larger and ''fatter'' than that under GR gravity (red dashed curve). While the blue caustic is thinner in the lateral direction than the red one. The reason is that the potential of plasma lensing has a slightly different ellipticity than that of gravitational lensing (the potential of plasma lensing is proportional to the electron density). The lensed images are closer to the lens with plasma than the GR case.
For lensing with exponential plasma model (right panel in Fig.\,\ref{fig:SIE mocked FRB}), we have $A=10^{-12}$, $B_{exp} = 0.0001$. The lensed image positions produced by the two theories are indistinguishable. In additional tests with higher electron density, we find more noticeable changes in image positions and shape of caustics as well.
When the source located near the caustics, the tiny difference between $f(T)$ and GR can be manifested, from the number of images, the image positions and the magnifications.
%
%

In this work, we assume that the electron density distribution is the same as the mass distribution, thus the plasma lensing effect only causes changes in the image position along the radial directions. While $f(T)$ gravity influences the image position in both radial and tangential directions. The changes induced by $f(T)$ gravity mainly depend on the ellipticity of the mass distribution $f$ as well as $A$ (see appendix for more details).

\begin{center}
\begin{figure*}  
  {\mbox{\includegraphics[width=3.3 in]{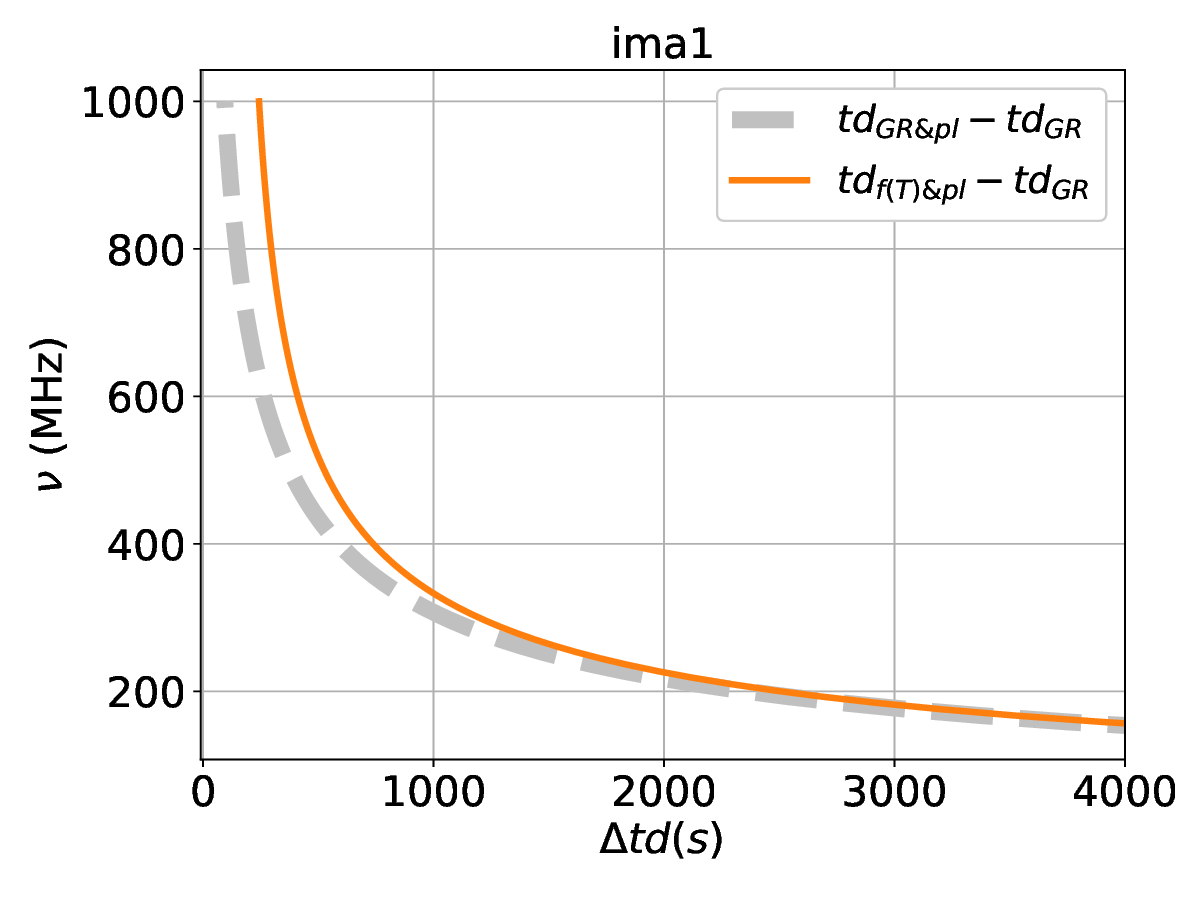}}}{\includegraphics[width=3.3 in]{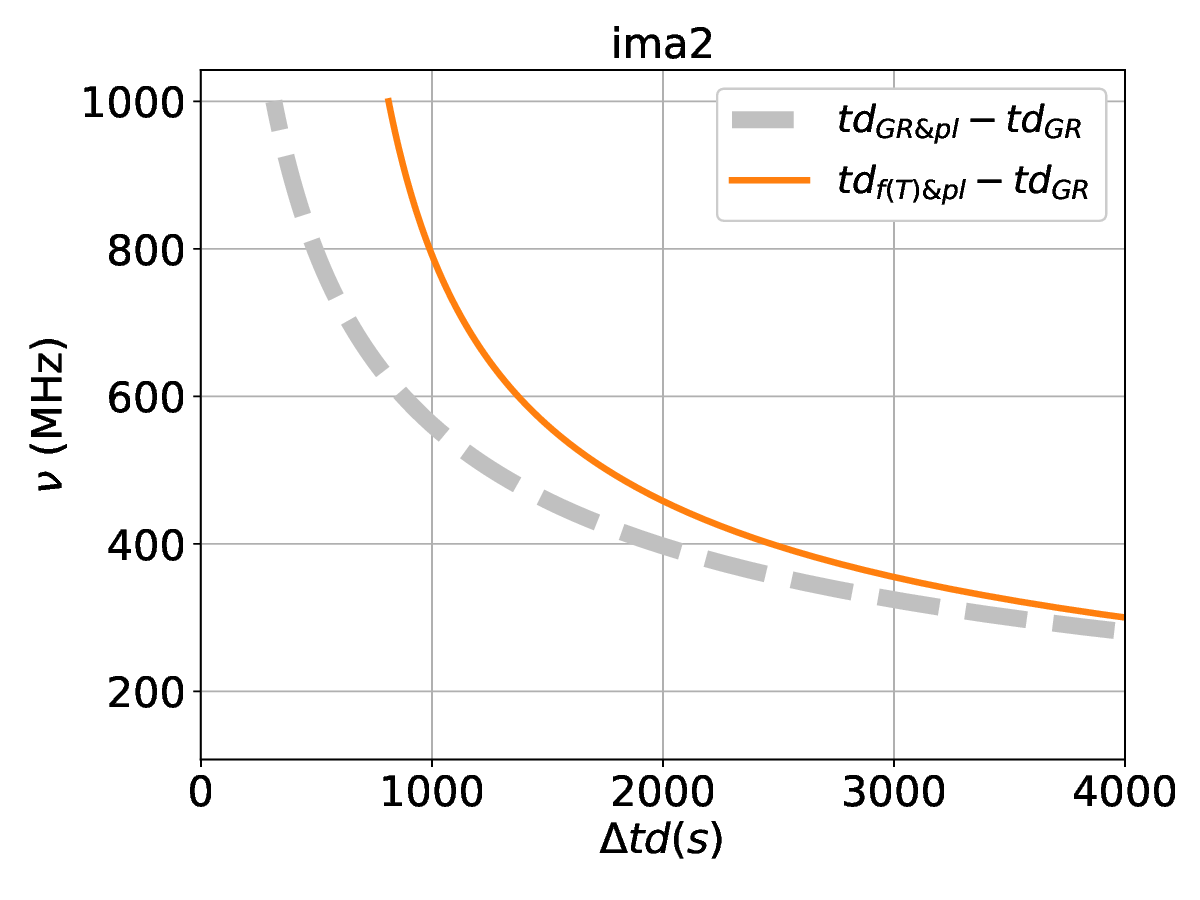}}
  \caption{The frequency-delay time relation for the two lensed images in len1 system. The delay time (in unit of second) caused by GR is subtracted for better comparison. An extremely high density is used when producing the figure. 
  }
\label{fig:fobs timedelay}
\end{figure*}
\end{center}

Although the image positions can be precisely measured, e.g. from VLBI, it is difficult to distinguish the gravity theory from that only, especially due to the degeneracy between plasma and gravity. The magnification can be contaminated by several factors, such as subhaloes, thus is difficult to be calibrated either. 
However, it's worth noting that time measurements can be achieved with high precision \cite[e.g.][]{Deller2008ExtremelyHP,Desvignes2016HighprecisionTO,Goldstein2017PreciseTD}. 
In Table\,\ref{tab:SIE_mock plasma lensing}, we display the results obtained from our mock lensed FRB with power-law plasma, of which the corresponding image positions are shown in the left column of Fig.\,\ref{fig:SIE mocked FRB}. In all three examples, we tune the lens parameters ($A$, $B_{pl}$ and source position), in order to obtain the similar or even the same image positions between GR and $f(T)$. 
From Table.\,\ref{tab:SIE_mock plasma lensing}, we can see that for three pairs of lensing systems, the lensed image positions from both GR and $f(T)$ lens are the same. The magnifications of the images differ among each pair. Especially those near the critical curve, there exists large difference. More importantly, the time delay between multiple images exhibits differences from tens of seconds to hundreds of seconds, which can be easily distinguished by current facilities. In Fig.\,\ref{fig:fobs timedelay} we present the frequency-delay relation of the two lensed images in $Len1$ system (top left panel in Fig.\,\ref{fig:SIE mocked FRB}). For simplicity, we use the image position at $375$ MHz, and calculate the arrival time of the signal, i.e. the change of image position due to frequency is not taken into account. In all cases, the extra delay in Fig.\,\ref{fig:fobs timedelay} are positive, and the delay caused by $f(T)$ is higher than that by GR. The two curves in each panel show different slopes which can be used to distinguish the gravity. For the image 2, which is closer to the lens, the difference in arrival time is larger. For both two images, the green and orange solid and dashed curves converge at low frequencies, where the plasma lensing dominates. Thus, it cannot provide good constraint to the gravity theory. Comparatively, the best frequency range to distinguish the $f(T)$ gravity will be around $400$ MHz, where the change of slope is significant. In our simulations, we use extremely high electron density to exaggerate the plasma lensing effects, e.g. the electron density near the image position is $\sim10^5 - 10^6$ pc\,cm$^{-3}$. In a more realistic case, $\sim10$ pc\,cm$^{-3}$, the extra time delay is small, $\sim1$ second, but the offset between the solid and dashed curve does not change much since that is caused by the $f(T)$ gravity.

\begin{table*}
\begin{center}
\begin{tabular}{ccccccc}
\hline
\hline 
\noalign{\smallskip}
\textbf{Lens} & \textbf{$A \& B_{pl}$} & \textbf{image} & \textbf{(x1,x2)} & \textbf{(y1,y2)} 
 & \textbf{Magnification} & \textbf{timedelay(s)}\\ 
\noalign{\smallskip}
\hline
\hline
\noalign{\smallskip}
\textbf{Len1}& $A=0$               & $ima1$ & $(+1.62, 0)$ & {(0.55,0.0)} & {1.9} & {0}\\
{ }          & $B_{pl} = 0.00873$  & $ima2$ & $(-0.48,0)$ & {  }          & {1.3}& {64606} \\

\noalign{\smallskip}
\hline
\noalign{\smallskip}

{  }& $A=0.001$       & $ima1$ & $(+1.62, 0)$ & {(0.55,0.0)} & {1.9} & {0}\\
{ } & $B_{pl} = 0.011$& $ima2$ & $(-0.48,0)$ & {  }          & {1.1} & {65300} \\

\noalign{\smallskip}
\hline 
\hline
\noalign{\smallskip}

\textbf{Len2} & $A = 0 $ & $ima1$ & $(+0.64, +0.80)$  &{(-0.1,0.1815)} &{46.0}  & {0}\\
{ }   & ${ B_{pl} = 0.00926}$& $ima2$ & $(+0.47, +0.91)$  & { }            &{9.1}   & {3885}\\
{ }   & ${ }$            & $ima3$ & $(-1.10, +0.48)$  & { }            &{3.0}   & {-10861}\\
{ }   & ${ }$            & $ima4$ & $(+0.10, -0.64)$  & { }            &{0.8}   & {18489}\\

\noalign{\smallskip}
\hline
\noalign{\smallskip}

\textbf{  } & $A =0.001$ & $ima1$ & $(+0.64, +0.80)$  &{(-0.1,0.18)}  & {89.8} & {0}\\
{ }   & ${ B_{pl} = 0.011}$  & $ima2$ & $(+0.47, +0.91)$  & { }           & {8.5} & {3779}\\
{ }   & ${ }$            & $ima3$ & $(-1.10, +0.48)$ & { }            & {3.1}  & {-10995}\\
{ }   & ${ }$            & $ima4$ & $(+0.10, -0.64)$ & { }            & {0.8}  & {18846}\\

\noalign{\smallskip}
\hline 
\hline
\noalign{\smallskip}

\textbf{Len3} & $A = 0$ & $ima1$ &          $(0.18, 1.32)$  & {(0.0,0.4615)} & {71.3} & {0}\\
{ }   & ${ B_{pl} = 0.01034}$   & $ima2$ & $(0, 1.33)$     & { }            & {28.5} & {575}\\
{ }   & ${ }$                    & $ima3$ & $(-0.18,1.32)$  & { }            & {71.3} & {0}\\
{ }   & ${ }$                    & $ima4$ & $(0,-0.33)$     & { }            & {0.19} & {54160}\\

\noalign{\smallskip}
\hline
\noalign{\smallskip}
\textbf{}   & $A = 0.001$     & $ima1$ & $(0.18, 1.32)$  & {(0.0,0.46  )} & {59.6} & {0}\\
{ }   & ${ B_{pl} = 0.011}$   & $ima2$ & $(0, 1.33)$     & { }            & {26.8} & {562}\\
{ }   & ${ }$                 & $ima3$ & $(-0.18, 1.32)$ & { }            & {59.6} & {0}\\
{ }   & ${ }$                 & $ima4$ & $(0,-0.33)$     & { }            & {0.17} & {54298}\\

\noalign{\smallskip}
\hline
\hline
\noalign{\smallskip}

\end{tabular}
\end{center}
\caption{The properties of lensed images. Len1, Len2, and Len3 correspond to the three panels in the left column of Fig.\,\ref{fig:SIE mocked FRB}, from top to bottom. In each lens two sets of parameters ($A\&B_{pl}$) are adopted for comparison. Those with $A=0$ corresponds to GR. $(y_1,y_2)-$source position, $(x_1,x_2)-$image position. The last column is the time delay between multiple images: $\Delta t= t_i- t_1$, where $t_i$ is the arrival time of image $i$.}
\label{tab:SIE_mock plasma lensing}
\end{table*}

\section{summary}
\label{sect:summary}
In this study, we investigate the gravitational lensing effects within the framework of $f(T)$ gravity, in particular for the SIS and SIE mass models. Our findings reveal that in $f(T)$ gravity, for both mass models, the deflection angle is greater than that in GR. This phenomenon becomes significant when the source and lens are well aligned. We show the parameter region of possible detection of image positions for the SIS model as shown in Fig.\,\ref{fig:SIS_alpha(x)}. 
In the gravitational lensing effect, the term "Cut" refers to the position where the number of images changes due to the changes in source position. In the SIS model, GR generates either one or two images. Whereas in $f(T)$ gravity, there are always two images. As depicted in Fig.\,\ref{fig:SIS_xy}, when the source lies outside the GR's "Cut", $f(T)$ gravity generates one image close to the result of GR while the other is close to the lens center and de-magnified. Similarly, in the SIE model, the "Cut" is also absent. And since the change in the number of images is influenced by the position of the source with respect to the critical curve and caustic, the specific result depends on the parameter of $f(T)$ gravity, $\bm{\mathring{\alpha}}$. For both SIS and SIE models, the critical curve generated by $f(T)$ gravity will be larger than that by GR, and the exact distortion depends on the value of $\bm{\mathring{\alpha}}$.

In most cases, the difference between GR and $f(T)$ is tiny. Although radio facilities allow for precise measurements of image positions, the degeneracy between modified gravitational lensing and plasma lensing hinders us to differentiate $f(T)$ theory from GR. Other lensing signatures, such as magnification and time delay, are essential. FRBs are bright radio sources with milli-seconds duration and are able to provide unprecedented high precision in the time delay measurement. However, since ionized interstellar medium exists generally in galaxies, the plasma lensing effects cannot be neglected when we study the lensed FRBs. We perform toy simulations of lensed FRBs including both gravitational and plasma lensing, and compare the lensing effects between GR and $f(T)$ theory. Plasma lensing can cause negative deflection angle and change all the lensing effects. In some particular cases, e.g., axis-symmetric model and some specific radial profiles, the plasma can compensate the change by $f(T)$ gravity (see Eq.\,\ref{eq:SIS_alpha_fTpl} and Fig.\,\ref{fig:alpha_fTexp} for SIS model and Table.\,\ref{tab:SIE_mock plasma lensing} for SIE model). But in general, the difference caused by $f(T)$ gravity shows distinguish behaviours. 
In our simulations, we specifically select our lensing properties and source position to have similar or even exact same image positions produced by GR and $f(T)$ gravity for three different cases: source is on the horizontal axis, near the fold and inside the cusp. In all three cases, we find significant differences in magnification and time delays. 
In additional tests with different plasma models, similar discrepancies can be found. Therefore, we expect that there is a possibility to examine the gravity theory through strongly lensed FRBs, even with the influence of plasma, although accurate lens modeling and time delay measurement are necessary.

In summary, our research centered on the analysis of gravitational lensing effects within the framework of $f(T)$ gravity, taking the additional influence of plasma into account, particularly in the context of lensed FRBs. Our findings suggest that it is possible to test and differentiate gravitational theories through strongly lensed FRBs. 
However, our models are far more simple in several aspects. For instance, we only consider the macro lensing model. The substructures, and more importantly the microlensing effects are not taken into account \cite[e.g.][]{Chen2021FRBsLB}. Given the small physical size of FRBs, the microlensing effects are expected to be large and non-negligible. The free electrons along the light of sight, e.g. in the Milky Way, host galaxy, or intergalactic medium, can cause perturbations as well, thus the real observational data will contain more structures and lead to further degeneracy or bias. Even the lens model of the whole galaxy, large uncertainty still exists. Some of observed signatures that discussed in this work can be attributed to either $f(T)$ gravity or the mass profile of the lens, e.g. a slightly different power index. Additionally, in the context of $f(T)$ gravity, the ''isothermal'' halo profile can diverge from a power law with index of $-2$, and that is beyond the scope of this work. On top of that, so far there is not a single strongly lensed FRB that has been confirmed. How to search for and identify the lensed FRB events will be a more interesting and important question at the moment. 
On the other hand, in order to test gravity, a strong field will be a better laboratory, e.g. near black hole \cite[e.g.][]{Perlick2021CalculatingBH}. Thus the shadow of a black hole can provide even stringent constraints to the gravity theory.

\section*{Data Availability}
The data underlying this article will be shared on reasonable request to the corresponding author.

\section*{Acknowledgements}
We would like to thank the referee Prof. Marek Biesiada for valuable comments on the draft. We thank Prof. Alessandro Sonnenfeld, Emmanuel Saridakis and Yaqi Zhao for their helpful discussions. 
XE is supported by the NSFC Grant No. 11933002, and the China Manned Space Project with No.CMS-CSST-2021-A07, No.CMS-CSST-2021-A12. Y.F.C. is supported in part by National Key R\&D Program of China (2021YFC2203100), by CAS Young Interdisciplinary Innovation Team (JCTD-2022-20), by NSFC (12261131497), by 111 Project for ''Observational and Theoretical Research on Dark Matter and Dark Energy'' (B23042), by Fundamental Research Funds for Central Universities, by CSC Innovation Talent Funds, by USTC Fellowship for International Cooperation, by USTC Research Funds of the Double First-Class Initiative.

\bibliographystyle{mnras}
\bibliography{ref}

\begin{thebibliography}{}
\makeatletter
\relax
\def\mn@urlcharsother{\let\do\@makeother \do\$\do\&\do\#\do\^\do\_\do\%\do\~}
\def\mn@doi{\begingroup\mn@urlcharsother \@ifnextchar [ {\mn@doi@}
  {\mn@doi@[]}}
\def\mn@doi@[#1]#2{\def\@tempa{#1}\ifx\@tempa\@empty \href
  {http://dx.doi.org/#2} {doi:#2}\else \href {http://dx.doi.org/#2} {#1}\fi
  \endgroup}
\def\mn@eprint#1#2{\mn@eprint@#1:#2::\@nil}
\def\mn@eprint@arXiv#1{\href {http://arxiv.org/abs/#1} {{\tt arXiv:#1}}}
\def\mn@eprint@dblp#1{\href {http://dblp.uni-trier.de/rec/bibtex/#1.xml}
  {dblp:#1}}
\def\mn@eprint@#1:#2:#3:#4\@nil{\def\@tempa {#1}\def\@tempb {#2}\def\@tempc
  {#3}\ifx \@tempc \@empty \let \@tempc \@tempb \let \@tempb \@tempa \fi \ifx
  \@tempb \@empty \def\@tempb {arXiv}\fi \@ifundefined
  {mn@eprint@\@tempb}{\@tempb:\@tempc}{\expandafter \expandafter \csname
  mn@eprint@\@tempb\endcsname \expandafter{\@tempc}}}

\bibitem[\protect\citeauthoryear{{A Surdej, J.}, {Fraipont-Caro, D.}, {Gosset,
  E.}  \& {Remy, M.}}{{A Surdej, J.} et~al.}{1993}]{Surdej1993GravitationalLI}
{A Surdej, J.} {Fraipont-Caro, D.} {Gosset, E.} R.,   {Remy, M.} eds, 1993,
  {Gravitational lenses in the universe}  Liege International Astrophysical
  Colloquia Vol. 31

\bibitem[\protect\citeauthoryear{{Abadi} \& {Kovetz}}{{Abadi} \&
  {Kovetz}}{2021}]{Abadi2021ProbingGS}
{Abadi} T.,  {Kovetz} E.~D.,  2021, \mn@doi [\prd]
  {10.1103/PhysRevD.104.103515}, \href
  {https://ui.adsabs.harvard.edu/abs/2021PhRvD.104j3515A} {104, 103515}

\bibitem[\protect\citeauthoryear{{Afrin}, {Vagnozzi}  \& {Ghosh}}{{Afrin}
  et~al.}{2023}]{Afrin2022TestsOL}
{Afrin} M.,  {Vagnozzi} S.,   {Ghosh} S.~G.,  2023, \mn@doi [\apj]
  {10.3847/1538-4357/acb334}, \href
  {https://ui.adsabs.harvard.edu/abs/2023ApJ...944..149A} {944, 149}

\bibitem[\protect\citeauthoryear{Alhamzawi \& Alhamzawi}{Alhamzawi \&
  Alhamzawi}{2016a}]{Alhamzawi2016GravitationalLB}
Alhamzawi A.,  Alhamzawi R.,  2016a, International Journal of Modern Physics D,
  25, 1650020

\bibitem[\protect\citeauthoryear{Alhamzawi \& Alhamzawi}{Alhamzawi \&
  Alhamzawi}{2016b}]{Alhamzawi2016GravitationalLI}
Alhamzawi A.,  Alhamzawi R.,  2016b, General Relativity and Gravitation, 48, 1

\bibitem[\protect\citeauthoryear{{Ayzenberg} \& {Yunes}}{{Ayzenberg} \&
  {Yunes}}{2018}]{Ayzenberg2018BlackHS}
{Ayzenberg} D.,  {Yunes} N.,  2018, \mn@doi [Classical and Quantum Gravity]
  {10.1088/1361-6382/aae87b}, \href
  {https://ui.adsabs.harvard.edu/abs/2018CQGra..35w5002A} {35, 235002}

\bibitem[\protect\citeauthoryear{{Bahamonde}, {Levi Said}  \&
  {Zubair}}{{Bahamonde} et~al.}{2020}]{Bahamonde2020SolarST}
{Bahamonde} S.,  {Levi Said} J.,   {Zubair} M.,  2020, \mn@doi [\jcap]
  {10.1088/1475-7516/2020/10/024}, \href
  {https://ui.adsabs.harvard.edu/abs/2020JCAP...10..024B} {2020, 024}

\bibitem[\protect\citeauthoryear{{Bahamonde} et~al.,}{{Bahamonde}
  et~al.}{2023}]{Bahamonde2021TeleparallelGF}
{Bahamonde} S.,  et~al., 2023, \mn@doi [Reports on Progress in Physics]
  {10.1088/1361-6633/ac9cef}, \href
  {https://ui.adsabs.harvard.edu/abs/2023RPPh...86b6901B} {86, 026901}

\bibitem[\protect\citeauthoryear{{Bamba}, {Geng}, {Lee}  \& {Luo}}{{Bamba}
  et~al.}{2011}]{Bamba2010EquationOS}
{Bamba} K.,  {Geng} C.-Q.,  {Lee} C.-C.,   {Luo} L.-W.,  2011, \mn@doi [\jcap]
  {10.1088/1475-7516/2011/01/021}, \href
  {https://ui.adsabs.harvard.edu/abs/2011JCAP...01..021B} {2011, 021}

\bibitem[\protect\citeauthoryear{{Bamba}, {Odintsov}  \&
  {S{\'a}ez-G{\'o}mez}}{{Bamba} et~al.}{2013}]{Bamba2013ConformalSA}
{Bamba} K.,  {Odintsov} S.~D.,   {S{\'a}ez-G{\'o}mez} D.,  2013, \mn@doi [\prd]
  {10.1103/PhysRevD.88.084042}, \href
  {https://ui.adsabs.harvard.edu/abs/2013PhRvD..88h4042B} {88, 084042}

\bibitem[\protect\citeauthoryear{{Bengochea} \& {Ferraro}}{{Bengochea} \&
  {Ferraro}}{2009}]{Bengochea2008DarkTA}
{Bengochea} G.~R.,  {Ferraro} R.,  2009, \mn@doi [\prd]
  {10.1103/PhysRevD.79.124019}, \href
  {https://ui.adsabs.harvard.edu/abs/2009PhRvD..79l4019B} {79, 124019}

\bibitem[\protect\citeauthoryear{{Berti} et~al.,}{{Berti}
  et~al.}{2015}]{Berti2015TestingGR}
{Berti} E.,  et~al., 2015, \mn@doi [Classical and Quantum Gravity]
  {10.1088/0264-9381/32/24/243001}, \href
  {https://ui.adsabs.harvard.edu/abs/2015CQGra..32x3001B} {32, 243001}

\bibitem[\protect\citeauthoryear{{Binney} \& {Tremaine}}{{Binney} \&
  {Tremaine}}{1987}]{1987gady.book.....B}
{Binney} J.,  {Tremaine} S.,  1987, {Galactic dynamics}.
Princeton University Press

\bibitem[\protect\citeauthoryear{{Binney} \& {Tremaine}}{{Binney} \&
  {Tremaine}}{2008}]{Binney2008GalacticDS}
{Binney} J.,  {Tremaine} S.,  2008, {Galactic Dynamics: Second Edition}

\bibitem[\protect\citeauthoryear{{Bisnovatyi-Kogan} \&
  {Tsupko}}{{Bisnovatyi-Kogan} \&
  {Tsupko}}{2010}]{BisnovatyiKogan2010GravitationalLI}
{Bisnovatyi-Kogan} G.~S.,  {Tsupko} O.~Y.,  2010, \mn@doi [\mnras]
  {10.1111/j.1365-2966.2010.16290.x}, \href
  {https://ui.adsabs.harvard.edu/abs/2010MNRAS.404.1790B} {404, 1790}

\bibitem[\protect\citeauthoryear{{Bisnovatyi-Kogan} \&
  {Tsupko}}{{Bisnovatyi-Kogan} \&
  {Tsupko}}{2015}]{BisnovatyiKogan2015GravitationalLI}
{Bisnovatyi-Kogan} G.~S.,  {Tsupko} O.~Y.,  2015, \mn@doi [Plasma Physics
  Reports] {10.1134/S1063780X15070016}, \href
  {https://ui.adsabs.harvard.edu/abs/2015PlPhR..41..562B} {41, 562}

\bibitem[\protect\citeauthoryear{{Blandford} \& {Narayan}}{{Blandford} \&
  {Narayan}}{1986}]{Blandford1986FermatsPC}
{Blandford} R.,  {Narayan} R.,  1986, \mn@doi [\apj] {10.1086/164709}, \href
  {https://ui.adsabs.harvard.edu/abs/1986ApJ...310..568B} {310, 568}

\bibitem[\protect\citeauthoryear{{Blandford} \& {Narayan}}{{Blandford} \&
  {Narayan}}{1992}]{Blandford1992CosmologicalAO}
{Blandford} R.~D.,  {Narayan} R.,  1992, \mn@doi [\araa]
  {10.1146/annurev.astro.30.1.311}, \href
  {https://ui.adsabs.harvard.edu/abs/1992ARA&A..30..311B} {30, 311}

\bibitem[\protect\citeauthoryear{{B{\"o}hmer}, {Mussa}  \&
  {Tamanini}}{{B{\"o}hmer} et~al.}{2011}]{Bhmer2011ExistenceOR}
{B{\"o}hmer} C.~G.,  {Mussa} A.,   {Tamanini} N.,  2011, \mn@doi [Classical and
  Quantum Gravity] {10.1088/0264-9381/28/24/245020}, \href
  {https://ui.adsabs.harvard.edu/abs/2011CQGra..28x5020B} {28, 245020}

\bibitem[\protect\citeauthoryear{{Bozza} \& {Mancini}}{{Bozza} \&
  {Mancini}}{2004}]{Bozza2003TimeDI}
{Bozza} V.,  {Mancini} L.,  2004, \mn@doi [General Relativity and Gravitation]
  {10.1023/B:GERG.0000010486.58026.4f}, \href
  {https://ui.adsabs.harvard.edu/abs/2004GReGr..36..435B} {36, 435}

\bibitem[\protect\citeauthoryear{{Cai}, {Chen}, {Dent}, {Dutta}  \&
  {Saridakis}}{{Cai} et~al.}{2011}]{Cai2011MatterBC}
{Cai} Y.-F.,  {Chen} S.-H.,  {Dent} J.~B.,  {Dutta} S.,   {Saridakis} E.~N.,
  2011, \mn@doi [Classical and Quantum Gravity]
  {10.1088/0264-9381/28/21/215011}, \href
  {https://ui.adsabs.harvard.edu/abs/2011CQGra..28u5011C} {28, 215011}

\bibitem[\protect\citeauthoryear{{Cai}, {Capozziello}, {De Laurentis}  \&
  {Saridakis}}{{Cai} et~al.}{2016}]{Cai2015fTTG}
{Cai} Y.-F.,  {Capozziello} S.,  {De Laurentis} M.,   {Saridakis} E.~N.,  2016,
  \mn@doi [Reports on Progress in Physics] {10.1088/0034-4885/79/10/106901},
  \href {https://ui.adsabs.harvard.edu/abs/2016RPPh...79j6901C} {79, 106901}

\bibitem[\protect\citeauthoryear{{Caldwell}, {Dave}  \&
  {Steinhardt}}{{Caldwell} et~al.}{1998a}]{Riess1998ObservationalEF}
{Caldwell} R.~R.,  {Dave} R.,   {Steinhardt} P.~J.,  1998a, \mn@doi [\prl]
  {10.1103/PhysRevLett.80.1582}, \href
  {https://ui.adsabs.harvard.edu/abs/1998PhRvL..80.1582C} {80, 1582}

\bibitem[\protect\citeauthoryear{{Caldwell}, {Dave}  \&
  {Steinhardt}}{{Caldwell} et~al.}{1998b}]{Caldwell1997CosmologicalIO}
{Caldwell} R.~R.,  {Dave} R.,   {Steinhardt} P.~J.,  1998b, \mn@doi [\prl]
  {10.1103/PhysRevLett.80.1582}, \href
  {https://ui.adsabs.harvard.edu/abs/1998PhRvL..80.1582C} {80, 1582}

\bibitem[\protect\citeauthoryear{{Campigotto}, {Diaferio}, {Hernandez}  \&
  {Fatibene}}{{Campigotto} et~al.}{2017}]{Campigotto2016StrongGL}
{Campigotto} M.~C.,  {Diaferio} A.,  {Hernandez} X.,   {Fatibene} L.,  2017,
  \mn@doi [\jcap] {10.1088/1475-7516/2017/06/057}, \href
  {https://ui.adsabs.harvard.edu/abs/2017JCAP...06..057C} {2017, 057}

\bibitem[\protect\citeauthoryear{{Capozziello} \& {de Laurentis}}{{Capozziello}
  \& {de Laurentis}}{2011}]{CAPOZZIELLO2011167}
{Capozziello} S.,  {de Laurentis} M.,  2011, \mn@doi [\physrep]
  {10.1016/j.physrep.2011.09.003}, \href
  {https://ui.adsabs.harvard.edu/abs/2011PhR...509..167C} {509, 167}

\bibitem[\protect\citeauthoryear{{Chen} et~al.,}{{Chen}
  et~al.}{2016}]{Chen2016SHARPI}
{Chen} G. C.~F.,  et~al., 2016, \mn@doi [\mnras] {10.1093/mnras/stw991}, \href
  {https://ui.adsabs.harvard.edu/abs/2016MNRAS.462.3457C} {462, 3457}

\bibitem[\protect\citeauthoryear{{Chen}, {Li}, {Shu}  \& {Cao}}{{Chen}
  et~al.}{2019}]{Chen2018AssessingTE}
{Chen} Y.,  {Li} R.,  {Shu} Y.,   {Cao} X.,  2019, \mn@doi [\mnras]
  {10.1093/mnras/stz1902}, \href
  {https://ui.adsabs.harvard.edu/abs/2019MNRAS.488.3745C} {488, 3745}

\bibitem[\protect\citeauthoryear{{Chen}, {Luo}, {Cai}  \& {Saridakis}}{{Chen}
  et~al.}{2020}]{Chen2020NewTO}
{Chen} Z.,  {Luo} W.,  {Cai} Y.-F.,   {Saridakis} E.~N.,  2020, \mn@doi [\prd]
  {10.1103/PhysRevD.102.104044}, \href
  {https://ui.adsabs.harvard.edu/abs/2020PhRvD.102j4044C} {102, 104044}

\bibitem[\protect\citeauthoryear{{Chen}, {Shu}, {Li}  \& {Zheng}}{{Chen}
  et~al.}{2021}]{Chen2021FRBsLB}
{Chen} X.,  {Shu} Y.,  {Li} G.,   {Zheng} W.,  2021, \mn@doi [\apj]
  {10.3847/1538-4357/ac2c76}, \href
  {https://ui.adsabs.harvard.edu/abs/2021ApJ...923..117C} {923, 117}

\bibitem[\protect\citeauthoryear{{Clegg}, {Fey}  \& {Lazio}}{{Clegg}
  et~al.}{1998}]{Clegg1997TheGP}
{Clegg} A.~W.,  {Fey} A.~L.,   {Lazio} T. J.~W.,  1998, \mn@doi [\apj]
  {10.1086/305344}, \href
  {https://ui.adsabs.harvard.edu/abs/1998ApJ...496..253C} {496, 253}

\bibitem[\protect\citeauthoryear{{Clifton}, {Ferreira}, {Padilla}  \&
  {Skordis}}{{Clifton} et~al.}{2012}]{Clifton2011ModifiedGA}
{Clifton} T.,  {Ferreira} P.~G.,  {Padilla} A.,   {Skordis} C.,  2012, \mn@doi
  [\physrep] {10.1016/j.physrep.2012.01.001}, \href
  {https://ui.adsabs.harvard.edu/abs/2012PhR...513....1C} {513, 1}

\bibitem[\protect\citeauthoryear{{Clowe}, {Brada{\v{c}}}, {Gonzalez},
  {Markevitch}, {Randall}, {Jones}  \& {Zaritsky}}{{Clowe}
  et~al.}{2006}]{Clowe2006ADE}
{Clowe} D.,  {Brada{\v{c}}} M.,  {Gonzalez} A.~H.,  {Markevitch} M.,  {Randall}
  S.~W.,  {Jones} C.,   {Zaritsky} D.,  2006, \mn@doi [\apjl] {10.1086/508162},
  \href {https://ui.adsabs.harvard.edu/abs/2006ApJ...648L.109C} {648, L109}

\bibitem[\protect\citeauthoryear{{Connor} \& {Ravi}}{{Connor} \&
  {Ravi}}{2023}]{2023MNRAS.521.4024C}
{Connor} L.,  {Ravi} V.,  2023, \mn@doi [\mnras] {10.1093/mnras/stad667}, \href
  {https://ui.adsabs.harvard.edu/abs/2023MNRAS.521.4024C} {521, 4024}

\bibitem[\protect\citeauthoryear{{Cordes} \& {Chatterjee}}{{Cordes} \&
  {Chatterjee}}{2019}]{Cordes2019FastRB}
{Cordes} J.~M.,  {Chatterjee} S.,  2019, \mn@doi [\araa]
  {10.1146/annurev-astro-091918-104501}, \href
  {https://ui.adsabs.harvard.edu/abs/2019ARA&A..57..417C} {57, 417}

\bibitem[\protect\citeauthoryear{{Cordes}, {Wasserman}, {Hessels}, {Lazio},
  {Chatterjee}  \& {Wharton}}{{Cordes} et~al.}{2017}]{Cordes2017LensingOF}
{Cordes} J.~M.,  {Wasserman} I.,  {Hessels} J.~W.~T.,  {Lazio} T.~J.~W.,
  {Chatterjee} S.,   {Wharton} R.~S.,  2017, \mn@doi [\apj]
  {10.3847/1538-4357/aa74da}, \href
  {https://ui.adsabs.harvard.edu/abs/2017ApJ...842...35C} {842, 35}

\bibitem[\protect\citeauthoryear{{Cunha} \& {Herdeiro}}{{Cunha} \&
  {Herdeiro}}{2018}]{Cunha2018ShadowsAS}
{Cunha} P. V.~P.,  {Herdeiro} C. A.~R.,  2018, \mn@doi [General Relativity and
  Gravitation] {10.1007/s10714-018-2361-9}, \href
  {https://ui.adsabs.harvard.edu/abs/2018GReGr..50...42C} {50, 42}

\bibitem[\protect\citeauthoryear{{Dai} \& {Lu}}{{Dai} \&
  {Lu}}{2017}]{Dai2017ProbingMO}
{Dai} L.,  {Lu} W.,  2017, \mn@doi [\apj] {10.3847/1538-4357/aa8873}, \href
  {https://ui.adsabs.harvard.edu/abs/2017ApJ...847...19D} {847, 19}

\bibitem[\protect\citeauthoryear{{De Felice} \& {Tsujikawa}}{{De Felice} \&
  {Tsujikawa}}{2010}]{DeFelice2010fRT}
{De Felice} A.,  {Tsujikawa} S.,  2010, \mn@doi [Living Reviews in Relativity]
  {10.12942/lrr-2010-3}, \href
  {https://ui.adsabs.harvard.edu/abs/2010LRR....13....3D} {13, 3}

\bibitem[\protect\citeauthoryear{{DeBenedictis} \& {Iliji{\'c}}}{{DeBenedictis}
  \& {Iliji{\'c}}}{2016}]{DeBenedictis2016SphericallySV}
{DeBenedictis} A.,  {Iliji{\'c}} S.,  2016, \mn@doi [\prd]
  {10.1103/PhysRevD.94.124025}, \href
  {https://ui.adsabs.harvard.edu/abs/2016PhRvD..94l4025D} {94, 124025}

\bibitem[\protect\citeauthoryear{{Deller}, {Verbiest}, {Tingay}  \&
  {Bailes}}{{Deller} et~al.}{2008}]{Deller2008ExtremelyHP}
{Deller} A.~T.,  {Verbiest} J.~P.~W.,  {Tingay} S.~J.,   {Bailes} M.,  2008,
  \mn@doi [\apjl] {10.1086/592401}, \href
  {https://ui.adsabs.harvard.edu/abs/2008ApJ...685L..67D} {685, L67}

\bibitem[\protect\citeauthoryear{{Desvignes} et~al.,}{{Desvignes}
  et~al.}{2016}]{Desvignes2016HighprecisionTO}
{Desvignes} G.,  et~al., 2016, \mn@doi [\mnras] {10.1093/mnras/stw483}, \href
  {https://ui.adsabs.harvard.edu/abs/2016MNRAS.458.3341D} {458, 3341}

\bibitem[\protect\citeauthoryear{{Enander} \& {M{\"o}rtsell}}{{Enander} \&
  {M{\"o}rtsell}}{2013}]{Enander2013StrongLC}
{Enander} J.,  {M{\"o}rtsell} E.,  2013, \mn@doi [Journal of High Energy
  Physics] {10.1007/JHEP10(2013)031}, \href
  {https://ui.adsabs.harvard.edu/abs/2013JHEP...10..031E} {2013, 31}

\bibitem[\protect\citeauthoryear{{Er} \& {Mao}}{{Er} \&
  {Mao}}{2022}]{Er2022TheEO}
{Er} X.,  {Mao} S.,  2022, \mn@doi [\mnras] {10.1093/mnras/stac2323}, \href
  {https://ui.adsabs.harvard.edu/abs/2022MNRAS.516.2218E} {516, 2218}

\bibitem[\protect\citeauthoryear{{Er} \& {Rogers}}{{Er} \&
  {Rogers}}{2018}]{Er2017TwoFO}
{Er} X.,  {Rogers} A.,  2018, \mn@doi [\mnras] {10.1093/mnras/stx3290}, \href
  {https://ui.adsabs.harvard.edu/abs/2018MNRAS.475..867E} {475, 867}

\bibitem[\protect\citeauthoryear{{Er} \& {Rogers}}{{Er} \&
  {Rogers}}{2019}]{Er2019TwoFO}
{Er} X.,  {Rogers} A.,  2019, \mn@doi [\mnras] {10.1093/mnras/stz2073}, \href
  {https://ui.adsabs.harvard.edu/abs/2019MNRAS.488.5651E} {488, 5651}

\bibitem[\protect\citeauthoryear{{Farrugia} \& {Said}}{{Farrugia} \&
  {Said}}{2016}]{Farrugia2016StabilityOT}
{Farrugia} G.,  {Said} J.~L.,  2016, \mn@doi [\prd]
  {10.1103/PhysRevD.94.124054}, \href
  {https://ui.adsabs.harvard.edu/abs/2016PhRvD..94l4054F} {94, 124054}

\bibitem[\protect\citeauthoryear{{Ferreira}}{{Ferreira}}{2019}]{Ferreira2019CosmologicalTO}
{Ferreira} P.~G.,  2019, \mn@doi [\araa] {10.1146/annurev-astro-091918-104423},
  \href {https://ui.adsabs.harvard.edu/abs/2019ARA&A..57..335F} {57, 335}

\bibitem[\protect\citeauthoryear{{Frieman}, {Turner}  \& {Huterer}}{{Frieman}
  et~al.}{2008}]{Frieman2008DarkEA}
{Frieman} J.~A.,  {Turner} M.~S.,   {Huterer} D.,  2008, \mn@doi [\araa]
  {10.1146/annurev.astro.46.060407.145243}, \href
  {https://ui.adsabs.harvard.edu/abs/2008ARA&A..46..385F} {46, 385}

\bibitem[\protect\citeauthoryear{{GRAVITY Collaboration} et~al.,}{{GRAVITY
  Collaboration} et~al.}{2018}]{Abuter2018DetectionOT}
{GRAVITY Collaboration} et~al., 2018, \mn@doi [\aap]
  {10.1051/0004-6361/201833718}, \href
  {https://ui.adsabs.harvard.edu/abs/2018A&A...615L..15G} {615, L15}

\bibitem[\protect\citeauthoryear{{Gao}, {Li}  \& {Gao}}{{Gao}
  et~al.}{2022}]{Gao2022ProspectsOS}
{Gao} R.,  {Li} Z.,   {Gao} H.,  2022, \mn@doi [\mnras]
  {10.1093/mnras/stac2270}, \href
  {https://ui.adsabs.harvard.edu/abs/2022MNRAS.516.1977G} {516, 1977}

\bibitem[\protect\citeauthoryear{{Goldstein}, {Nugent}, {Kasen}  \&
  {Collett}}{{Goldstein} et~al.}{2018}]{Goldstein2017PreciseTD}
{Goldstein} D.~A.,  {Nugent} P.~E.,  {Kasen} D.~N.,   {Collett} T.~E.,  2018,
  \mn@doi [\apj] {10.3847/1538-4357/aaa975}, \href
  {https://ui.adsabs.harvard.edu/abs/2018ApJ...855...22G} {855, 22}

\bibitem[\protect\citeauthoryear{{G{\"u}rkan}, {Jackson}, {Browne}, {Koopmans},
  {Fassnacht}  \& {Berciano Alba}}{{G{\"u}rkan} et~al.}{2010}]{Gurkan2010AMF}
{G{\"u}rkan} G.,  {Jackson} N.,  {Browne} I.~W.~A.,  {Koopmans} L.~V.~E.,
  {Fassnacht} C.~D.,   {Berciano Alba} A.,  2010, \mn@doi [arXiv e-prints]
  {10.48550/arXiv.1011.6604}, \href
  {https://ui.adsabs.harvard.edu/abs/2010arXiv1011.6604G} {p. arXiv:1011.6604}

\bibitem[\protect\citeauthoryear{{Guti{\'e}rrez} \& {Beckman}}{{Guti{\'e}rrez}
  \& {Beckman}}{2010}]{2010ApJ...710L..44G}
{Guti{\'e}rrez} L.,  {Beckman} J.~E.,  2010, \mn@doi [\apjl]
  {10.1088/2041-8205/710/1/L44}, \href
  {https://ui.adsabs.harvard.edu/abs/2010ApJ...710L..44G} {710, L44}

\bibitem[\protect\citeauthoryear{{Hees} et~al.,}{{Hees}
  et~al.}{2017}]{Hees2017TestingGR}
{Hees} A.,  et~al., 2017, \mn@doi [\prl] {10.1103/PhysRevLett.118.211101},
  \href {https://ui.adsabs.harvard.edu/abs/2017PhRvL.118u1101H} {118, 211101}

\bibitem[\protect\citeauthoryear{{Hildebrandt} et~al.,}{{Hildebrandt}
  et~al.}{2017}]{Hildebrandt2016KiDS450CP}
{Hildebrandt} H.,  et~al., 2017, \mn@doi [\mnras] {10.1093/mnras/stw2805},
  \href {https://ui.adsabs.harvard.edu/abs/2017MNRAS.465.1454H} {465, 1454}

\bibitem[\protect\citeauthoryear{{Hohmann}}{{Hohmann}}{2023}]{Hohmann2022TeleparallelG}
{Hohmann} M.,  2023, in {Pfeifer} C.,  {L{\"a}mmerzahl} C.,  eds, , Vol.~1017,
  Lecture Notes in Physics, Berlin Springer Verlag.
pp 145--198, \mn@doi{10.1007/978-3-031-31520-6_4}

\bibitem[\protect\citeauthoryear{{Hurtado}, {Casta{\~n}eda}  \&
  {Tejeiro}}{{Hurtado} et~al.}{2014}]{Hurtado2013GravitationalLB}
{Hurtado} R.,  {Casta{\~n}eda} L.,   {Tejeiro} J.~M.,  2014, \mn@doi
  [International Journal of Astronomy and Astrophysics]
  {10.4236/ijaa.2014.42028}, \href
  {https://ui.adsabs.harvard.edu/abs/2014IJAA....4..340H} {4, 340}

\bibitem[\protect\citeauthoryear{{Iliji{\'c}} \& {Sossich}}{{Iliji{\'c}} \&
  {Sossich}}{2018}]{Iliji2018CompactSI}
{Iliji{\'c}} S.,  {Sossich} M.,  2018, \mn@doi [\prd]
  {10.1103/PhysRevD.98.064047}, \href
  {https://ui.adsabs.harvard.edu/abs/2018PhRvD..98f4047I} {98, 064047}

\bibitem[\protect\citeauthoryear{{Iorio} \& {Saridakis}}{{Iorio} \&
  {Saridakis}}{2012}]{Iorio2012SolarSC}
{Iorio} L.,  {Saridakis} E.~N.,  2012, \mn@doi [\mnras]
  {10.1111/j.1365-2966.2012.21995.x}, \href
  {https://ui.adsabs.harvard.edu/abs/2012MNRAS.427.1555I} {427, 1555}

\bibitem[\protect\citeauthoryear{{Iorio}, {Radicella}  \& {Ruggiero}}{{Iorio}
  et~al.}{2015}]{Iorio2015ConstrainingFG}
{Iorio} L.,  {Radicella} N.,   {Ruggiero} M.~L.,  2015, \mn@doi [\jcap]
  {10.1088/1475-7516/2015/08/021}, \href
  {https://ui.adsabs.harvard.edu/abs/2015JCAP...08..021I} {2015, 021}

\bibitem[\protect\citeauthoryear{{Iorio}, {Ruggiero}, {Radicella}  \&
  {Saridakis}}{{Iorio} et~al.}{2016}]{Iorio2016ConstrainingTS}
{Iorio} L.,  {Ruggiero} M.~L.,  {Radicella} N.,   {Saridakis} E.~N.,  2016,
  \mn@doi [Physics of the Dark Universe] {10.1016/j.dark.2016.05.001}, \href
  {https://ui.adsabs.harvard.edu/abs/2016PDU....13..111I} {13, 111}

\bibitem[\protect\citeauthoryear{{Johannsen}, {Wang}, {Broderick}, {Doeleman},
  {Fish}, {Loeb}  \& {Psaltis}}{{Johannsen}
  et~al.}{2016}]{Johannsen2016TestingGR}
{Johannsen} T.,  {Wang} C.,  {Broderick} A.~E.,  {Doeleman} S.~S.,  {Fish}
  V.~L.,  {Loeb} A.,   {Psaltis} D.,  2016, \mn@doi [\prl]
  {10.1103/PhysRevLett.117.091101}, \href
  {https://ui.adsabs.harvard.edu/abs/2016PhRvL.117i1101J} {117, 091101}

\bibitem[\protect\citeauthoryear{{Kassiola} \& {Kovner}}{{Kassiola} \&
  {Kovner}}{1993}]{Kassiola1993EllipticMD}
{Kassiola} A.,  {Kovner} I.,  1993, \mn@doi [\apj] {10.1086/173325}, \href
  {https://ui.adsabs.harvard.edu/abs/1993ApJ...417..450K} {417, 450}

\bibitem[\protect\citeauthoryear{{Keeton}}{{Keeton}}{2001}]{Keeton2001ACO}
{Keeton} C.~R.,  2001, \mn@doi [arXiv e-prints]
  {10.48550/arXiv.astro-ph/0102341}, \href
  {https://ui.adsabs.harvard.edu/abs/2001astro.ph..2341K} {pp
  astro--ph/0102341}

\bibitem[\protect\citeauthoryear{{Kelly} et~al.,}{{Kelly}
  et~al.}{2015}]{Kelly2014MultipleIO}
{Kelly} P.~L.,  et~al., 2015, \mn@doi [Science] {10.1126/science.aaa3350},
  \href {https://ui.adsabs.harvard.edu/abs/2015Sci...347.1123K} {347, 1123}

\bibitem[\protect\citeauthoryear{{Kormann}, {Schneider}  \&
  {Bartelmann}}{{Kormann} et~al.}{1994}]{Kormann1994IsothermalEG}
{Kormann} R.,  {Schneider} P.,   {Bartelmann} M.,  1994, \aap, \href
  {https://ui.adsabs.harvard.edu/abs/1994A&A...284..285K} {284, 285}

\bibitem[\protect\citeauthoryear{{Kravtsov} \& {Borgani}}{{Kravtsov} \&
  {Borgani}}{2012}]{Kravtsov2012FormationOG}
{Kravtsov} A.~V.,  {Borgani} S.,  2012, \mn@doi [\araa]
  {10.1146/annurev-astro-081811-125502}, \href
  {https://ui.adsabs.harvard.edu/abs/2012ARA&A..50..353K} {50, 353}

\bibitem[\protect\citeauthoryear{{Kr{\v{s}}{\v{s}}{\'a}k}, {van den Hoogen},
  {Pereira}, {B{\"o}hmer}  \& {Coley}}{{Kr{\v{s}}{\v{s}}{\'a}k}
  et~al.}{2019}]{Krk2018TeleparallelTO}
{Kr{\v{s}}{\v{s}}{\'a}k} M.,  {van den Hoogen} R.~J.,  {Pereira} J.~G.,
  {B{\"o}hmer} C.~G.,   {Coley} A.~A.,  2019, \mn@doi [Classical and Quantum
  Gravity] {10.1088/1361-6382/ab2e1f}, \href
  {https://ui.adsabs.harvard.edu/abs/2019CQGra..36r3001K} {36, 183001}

\bibitem[\protect\citeauthoryear{{Kuang}, {Tang}, {Wang}  \& {Wang}}{{Kuang}
  et~al.}{2022}]{Kuang2022ConstrainingAM}
{Kuang} X.-M.,  {Tang} Z.-Y.,  {Wang} B.,   {Wang} A.,  2022, \mn@doi [\prd]
  {10.1103/PhysRevD.106.064012}, \href
  {https://ui.adsabs.harvard.edu/abs/2022PhRvD.106f4012K} {106, 064012}

\bibitem[\protect\citeauthoryear{{Kumar} \& {Beniamini}}{{Kumar} \&
  {Beniamini}}{2023}]{Kumar2022GravitationalLI}
{Kumar} P.,  {Beniamini} P.,  2023, \mn@doi [\mnras] {10.1093/mnras/stad160},
  \href {https://ui.adsabs.harvard.edu/abs/2023MNRAS.520..247K} {520, 247}

\bibitem[\protect\citeauthoryear{{Li}, {Gao}, {Ding}, {Wang}  \& {Zhang}}{{Li}
  et~al.}{2018a}]{Li2017StronglyLR}
{Li} Z.-X.,  {Gao} H.,  {Ding} X.-H.,  {Wang} G.-J.,   {Zhang} B.,  2018a,
  \mn@doi [Nature Communications] {10.1038/s41467-018-06303-0}, \href
  {https://ui.adsabs.harvard.edu/abs/2018NatCo...9.3833L} {9, 3833}

\bibitem[\protect\citeauthoryear{{Li}, {Cai}, {Cai}  \& {Saridakis}}{{Li}
  et~al.}{2018b}]{Li2018TheEF}
{Li} C.,  {Cai} Y.,  {Cai} Y.-F.,   {Saridakis} E.~N.,  2018b, \mn@doi [\jcap]
  {10.1088/1475-7516/2018/10/001}, \href
  {https://ui.adsabs.harvard.edu/abs/2018JCAP...10..001L} {2018, 001}

\bibitem[\protect\citeauthoryear{{Liao}, {Biesiada}  \& {Zhu}}{{Liao}
  et~al.}{2022}]{Liao2022StronglyLT}
{Liao} K.,  {Biesiada} M.,   {Zhu} Z.-H.,  2022, \mn@doi [Chinese Physics
  Letters] {10.1088/0256-307X/39/11/119801}, \href
  {https://ui.adsabs.harvard.edu/abs/2022ChPhL..39k9801L} {39, 119801}

\bibitem[\protect\citeauthoryear{{Lorimer}, {Bailes}, {McLaughlin}, {Narkevic}
  \& {Crawford}}{{Lorimer} et~al.}{2007}]{Lorimer2007ABM}
{Lorimer} D.~R.,  {Bailes} M.,  {McLaughlin} M.~A.,  {Narkevic} D.~J.,
  {Crawford} F.,  2007, \mn@doi [Science] {10.1126/science.1147532}, \href
  {https://ui.adsabs.harvard.edu/abs/2007Sci...318..777L} {318, 777}

\bibitem[\protect\citeauthoryear{{Maluf}}{{Maluf}}{2013}]{Maluf2013TheTE}
{Maluf} J.~W.,  2013, \mn@doi [Annalen der Physik] {10.1002/andp.201200272},
  \href {https://ui.adsabs.harvard.edu/abs/2013AnP...525..339M} {525, 339}

\bibitem[\protect\citeauthoryear{{Marcote} et~al.,}{{Marcote}
  et~al.}{2017}]{Marcote2017TheRF}
{Marcote} B.,  et~al., 2017, \mn@doi [\apjl] {10.3847/2041-8213/834/2/L8},
  \href {https://ui.adsabs.harvard.edu/abs/2017ApJ...834L...8M} {834, L8}

\bibitem[\protect\citeauthoryear{{Mathews} \& {Brighenti}}{{Mathews} \&
  {Brighenti}}{2003}]{2003ARA&A..41..191M}
{Mathews} W.~G.,  {Brighenti} F.,  2003, \mn@doi [\araa]
  {10.1146/annurev.astro.41.090401.094542}, \href
  {https://ui.adsabs.harvard.edu/abs/2003ARA&A..41..191M} {41, 191}

\bibitem[\protect\citeauthoryear{{Meneghetti}}{{Meneghetti}}{2022}]{Meneghetti2021IntroductionTG}
{Meneghetti} M.,  2022, {Introduction to Gravitational Lensing: With Python
  Examples}

\bibitem[\protect\citeauthoryear{{Mizuno} et~al.,}{{Mizuno}
  et~al.}{2018}]{Mizuno2018TheCA}
{Mizuno} Y.,  et~al., 2018, \mn@doi [Nature Astronomy]
  {10.1038/s41550-018-0449-5}, \href
  {https://ui.adsabs.harvard.edu/abs/2018NatAs...2..585M} {2, 585}

\bibitem[\protect\citeauthoryear{{Mu{\~n}oz}, {Kovetz}, {Dai}  \&
  {Kamionkowski}}{{Mu{\~n}oz} et~al.}{2016}]{Muoz2016LensingOF}
{Mu{\~n}oz} J.~B.,  {Kovetz} E.~D.,  {Dai} L.,   {Kamionkowski} M.,  2016,
  \mn@doi [\prl] {10.1103/PhysRevLett.117.091301}, \href
  {https://ui.adsabs.harvard.edu/abs/2016PhRvL.117i1301M} {117, 091301}

\bibitem[\protect\citeauthoryear{{Narayan} \& {Bartelmann}}{{Narayan} \&
  {Bartelmann}}{1996}]{Narayan1996LecturesOG}
{Narayan} R.,  {Bartelmann} M.,  1996, \mn@doi [arXiv e-prints]
  {10.48550/arXiv.astro-ph/9606001}, \href
  {https://ui.adsabs.harvard.edu/abs/1996astro.ph..6001N} {pp
  astro--ph/9606001}

\bibitem[\protect\citeauthoryear{{Narikawa}, {Kobayashi}, {Yamauchi}  \&
  {Saito}}{{Narikawa} et~al.}{2013}]{Narikawa2013TestingGS}
{Narikawa} T.,  {Kobayashi} T.,  {Yamauchi} D.,   {Saito} R.,  2013, \mn@doi
  [\prd] {10.1103/PhysRevD.87.124006}, \href
  {https://ui.adsabs.harvard.edu/abs/2013PhRvD..87l4006N} {87, 124006}

\bibitem[\protect\citeauthoryear{{Nazari}}{{Nazari}}{2022}]{Nazari2022LightBA}
{Nazari} E.,  2022, \mn@doi [\prd] {10.1103/PhysRevD.105.104026}, \href
  {https://ui.adsabs.harvard.edu/abs/2022PhRvD.105j4026N} {105, 104026}

\bibitem[\protect\citeauthoryear{{Nojiri} \& {Odintsov}}{{Nojiri} \&
  {Odintsov}}{2006}]{Nojiri2006MODIFIEDF}
{Nojiri} S.,  {Odintsov} S.~D.,  2006, \mn@doi [\prd]
  {10.1103/PhysRevD.74.086005}, \href
  {https://ui.adsabs.harvard.edu/abs/2006PhRvD..74h6005N} {74, 086005}

\bibitem[\protect\citeauthoryear{{Nzioki}, {Dunsby}, {Goswami}  \&
  {Carloni}}{{Nzioki} et~al.}{2011}]{Nzioki2010AGA}
{Nzioki} A.~M.,  {Dunsby} P. K.~S.,  {Goswami} R.,   {Carloni} S.,  2011,
  \mn@doi [\prd] {10.1103/PhysRevD.83.024030}, \href
  {https://ui.adsabs.harvard.edu/abs/2011PhRvD..83b4030N} {83, 024030}

\bibitem[\protect\citeauthoryear{{Oguri}}{{Oguri}}{2019}]{Oguri2019StrongGL}
{Oguri} M.,  2019, \mn@doi [Reports on Progress in Physics]
  {10.1088/1361-6633/ab4fc5}, \href
  {https://ui.adsabs.harvard.edu/abs/2019RPPh...82l6901O} {82, 126901}

\bibitem[\protect\citeauthoryear{{Oguri}, {Taruya}, {Suto}  \&
  {Turner}}{{Oguri} et~al.}{2002}]{Oguri2001StrongGL}
{Oguri} M.,  {Taruya} A.,  {Suto} Y.,   {Turner} E.~L.,  2002, \mn@doi [\apj]
  {10.1086/339064}, \href
  {https://ui.adsabs.harvard.edu/abs/2002ApJ...568..488O} {568, 488}

\bibitem[\protect\citeauthoryear{{Panpanich}, {Ponglertsakul}  \&
  {Tannukij}}{{Panpanich} et~al.}{2019}]{Panpanich2019ParticleMA}
{Panpanich} S.,  {Ponglertsakul} S.,   {Tannukij} L.,  2019, \mn@doi [\prd]
  {10.1103/PhysRevD.100.044031}, \href
  {https://ui.adsabs.harvard.edu/abs/2019PhRvD.100d4031P} {100, 044031}

\bibitem[\protect\citeauthoryear{{Perlick} \& {Tsupko}}{{Perlick} \&
  {Tsupko}}{2022}]{Perlick2021CalculatingBH}
{Perlick} V.,  {Tsupko} O.~Y.,  2022, \mn@doi [\physrep]
  {10.1016/j.physrep.2021.10.004}, \href
  {https://ui.adsabs.harvard.edu/abs/2022PhR...947....1P} {947, 1}

\bibitem[\protect\citeauthoryear{{Perlick} \& {Tsupko}}{{Perlick} \&
  {Tsupko}}{2023}]{2023arXiv231110615P}
{Perlick} V.,  {Tsupko} O.~Y.,  2023, \mn@doi [arXiv:2311.10615]
  {10.48550/arXiv.2311.10615}, \href
  {https://ui.adsabs.harvard.edu/abs/2023arXiv231110615P} {p. arXiv:2311.10615}

\bibitem[\protect\citeauthoryear{{Perlmutter} et~al.,}{{Perlmutter}
  et~al.}{1999}]{Perlmutter1998MeasurementsO}
{Perlmutter} S.,  et~al., 1999, \mn@doi [\apj] {10.1086/307221}, \href
  {https://ui.adsabs.harvard.edu/abs/1999ApJ...517..565P} {517, 565}

\bibitem[\protect\citeauthoryear{{Psaltis} et~al.,}{{Psaltis}
  et~al.}{2020}]{Psaltis2020GravitationalTB}
{Psaltis} D.,  et~al., 2020, \mn@doi [\prl] {10.1103/PhysRevLett.125.141104},
  \href {https://ui.adsabs.harvard.edu/abs/2020PhRvL.125n1104P} {125, 141104}

\bibitem[\protect\citeauthoryear{{Refsdal}}{{Refsdal}}{1964}]{Refsdal1964TheGL}
{Refsdal} S.,  1964, \mn@doi [\mnras] {10.1093/mnras/128.4.295}, \href
  {https://ui.adsabs.harvard.edu/abs/1964MNRAS.128..295R} {128, 295}

\bibitem[\protect\citeauthoryear{{Ren}, {Wong}, {Cai}  \& {Saridakis}}{{Ren}
  et~al.}{2021a}]{Ren2021DatadrivenRO}
{Ren} X.,  {Wong} T. H.~T.,  {Cai} Y.-F.,   {Saridakis} E.~N.,  2021a, \mn@doi
  [Physics of the Dark Universe] {10.1016/j.dark.2021.100812}, \href
  {https://ui.adsabs.harvard.edu/abs/2021PDU....3200812R} {32, 100812}

\bibitem[\protect\citeauthoryear{{Ren}, {Zhao}, {Saridakis}  \& {Cai}}{{Ren}
  et~al.}{2021b}]{Ren2021DeflectionAA}
{Ren} X.,  {Zhao} Y.,  {Saridakis} E.~N.,   {Cai} Y.-F.,  2021b, \mn@doi
  [\jcap] {10.1088/1475-7516/2021/10/062}, \href
  {https://ui.adsabs.harvard.edu/abs/2021JCAP...10..062R} {2021, 062}

\bibitem[\protect\citeauthoryear{{Ren}, {Yan}, {Zhao}, {Cai}  \&
  {Saridakis}}{{Ren} et~al.}{2022}]{Ren2022GaussianPA}
{Ren} X.,  {Yan} S.-F.,  {Zhao} Y.,  {Cai} Y.-F.,   {Saridakis} E.~N.,  2022,
  \mn@doi [\apj] {10.3847/1538-4357/ac6ba5}, \href
  {https://ui.adsabs.harvard.edu/abs/2022ApJ...932..131R} {932, 131}

\bibitem[\protect\citeauthoryear{{Ruggiero}}{{Ruggiero}}{2016}]{Ruggiero2016LightBI}
{Ruggiero} M.~L.,  2016, \mn@doi [International Journal of Modern Physics D]
  {10.1142/S0218271816500735}, \href
  {https://ui.adsabs.harvard.edu/abs/2016IJMPD..2550073R} {25, 1650073}

\bibitem[\protect\citeauthoryear{{Ruggiero} \& {Radicella}}{{Ruggiero} \&
  {Radicella}}{2015}]{Ruggiero2015WeakFieldSS}
{Ruggiero} M.~L.,  {Radicella} N.,  2015, \mn@doi [\prd]
  {10.1103/PhysRevD.91.104014}, \href
  {https://ui.adsabs.harvard.edu/abs/2015PhRvD..91j4014R} {91, 104014}

\bibitem[\protect\citeauthoryear{{Schmidt}}{{Schmidt}}{2008}]{Schmidt2008WeakLP}
{Schmidt} F.,  2008, \mn@doi [\prd] {10.1103/PhysRevD.78.043002}, \href
  {https://ui.adsabs.harvard.edu/abs/2008PhRvD..78d3002S} {78, 043002}

\bibitem[\protect\citeauthoryear{{Schneider}}{{Schneider}}{2019}]{Schneider2014GeneralizedMG}
{Schneider} P.,  2019, \mn@doi [\aap] {10.1051/0004-6361/201424881}, \href
  {https://ui.adsabs.harvard.edu/abs/2019A&A...624A..54S} {624, A54}

\bibitem[\protect\citeauthoryear{{Schneider}, {Ehlers}  \& {Falco}}{{Schneider}
  et~al.}{1992a}]{https://doi.org/10.1002/asna.2113140412}
{Schneider} P.,  {Ehlers} J.,   {Falco} E.~E.,  1992a, {Gravitational Lenses},
  \mn@doi{10.1007/978-3-662-03758-4.
}

\bibitem[\protect\citeauthoryear{Schneider, Ehlers  \& Falco}{Schneider
  et~al.}{1992b}]{Schneider1992GravitationalLA}
Schneider P.,  Ehlers J.,   Falco E.~E.,  1992b, Gravitational lenses as
  astrophysical tools.
Springer Berlin Heidelberg, Berlin, Heidelberg, pp 467--515,
  \mn@doi{10.1007/978-3-662-03758-4_13}

\bibitem[\protect\citeauthoryear{{Sharma}, {Yadav}  \& {Verma}}{{Sharma}
  et~al.}{2021}]{2021EPJC...81..109S}
{Sharma} V.~K.,  {Yadav} B.~K.,   {Verma} M.~M.,  2021, \mn@doi [European
  Physical Journal C] {10.1140/epjc/s10052-021-08908-0}, \href
  {https://ui.adsabs.harvard.edu/abs/2021EPJC...81..109S} {81, 109}

\bibitem[\protect\citeauthoryear{{Sharma}, {Harikumar}, {Grespan}, {Biesiada}
  \& {Manohar Verma}}{{Sharma} et~al.}{2023}]{2023arXiv231010346S}
{Sharma} V.~K.,  {Harikumar} S.,  {Grespan} M.,  {Biesiada} M.,   {Manohar
  Verma} M.,  2023, \mn@doi [arXiv:2310.10346] {10.48550/arXiv.2310.10346},
  \href {https://ui.adsabs.harvard.edu/abs/2023arXiv231010346S} {p.
  arXiv:2310.10346}

\bibitem[\protect\citeauthoryear{{Sluse}, {Chantry}, {Magain}, {Courbin}  \&
  {Meylan}}{{Sluse} et~al.}{2012}]{Courbin2004COSMOGRAILTC}
{Sluse} D.,  {Chantry} V.,  {Magain} P.,  {Courbin} F.,   {Meylan} G.,  2012,
  \mn@doi [\aap] {10.1051/0004-6361/201015844}, \href
  {https://ui.adsabs.harvard.edu/abs/2012A&A...538A..99S} {538, A99}

\bibitem[\protect\citeauthoryear{{Sobouti}}{{Sobouti}}{2007}]{2007A&A...464..921S}
{Sobouti} Y.,  2007, \mn@doi [\aap] {10.1051/0004-6361:20065188}, \href
  {https://ui.adsabs.harvard.edu/abs/2007A&A...464..921S} {464, 921}

\bibitem[\protect\citeauthoryear{{Spitler} et~al.,}{{Spitler}
  et~al.}{2016}]{Spitler2016ARF}
{Spitler} L.~G.,  et~al., 2016, \mn@doi [\nat] {10.1038/nature17168}, \href
  {https://ui.adsabs.harvard.edu/abs/2016Natur.531..202S} {531, 202}

\bibitem[\protect\citeauthoryear{{Suyu}}{{Suyu}}{2012}]{Suyu2012AccurateCF}
{Suyu} S.,  2012, {Accurate Cosmology from Gravitational Lens Time Delays}, HST
  Proposal ID 12889. Cycle 20

\bibitem[\protect\citeauthoryear{{Suyu} et~al.,}{{Suyu}
  et~al.}{2013}]{Suyu2012TWOAT}
{Suyu} S.~H.,  et~al., 2013, \mn@doi [\apj] {10.1088/0004-637X/766/2/70}, \href
  {https://ui.adsabs.harvard.edu/abs/2013ApJ...766...70S} {766, 70}

\bibitem[\protect\citeauthoryear{{Tewes} et~al.,}{{Tewes}
  et~al.}{2013}]{Tewes2012COSMOGRAILTC}
{Tewes} M.,  et~al., 2013, \mn@doi [\aap] {10.1051/0004-6361/201220352}, \href
  {https://ui.adsabs.harvard.edu/abs/2013A&A...556A..22T} {556, A22}

\bibitem[\protect\citeauthoryear{{Thompson}, {Moran}  \& {Swenson}}{{Thompson}
  et~al.}{2017}]{2017isra.book.....T}
{Thompson} A.~R.,  {Moran} J.~M.,   {Swenson} George~W. J.,  2017,
  {Interferometry and Synthesis in Radio Astronomy, 3rd Edition},
  \mn@doi{10.1007/978-3-319-44431-4.
}

\bibitem[\protect\citeauthoryear{{Treu}}{{Treu}}{2010}]{Treu2010StrongLB}
{Treu} T.,  2010, \mn@doi [\araa] {10.1146/annurev-astro-081309-130924}, \href
  {https://ui.adsabs.harvard.edu/abs/2010ARA&A..48...87T} {48, 87}

\bibitem[\protect\citeauthoryear{{Treu}, {Suyu}  \& {Marshall}}{{Treu}
  et~al.}{2022}]{Treu2022StrongLT}
{Treu} T.,  {Suyu} S.~H.,   {Marshall} P.~J.,  2022, \mn@doi [\aapr]
  {10.1007/s00159-022-00145-y}, \href
  {https://ui.adsabs.harvard.edu/abs/2022A&ARv..30....8T} {30, 8}

\bibitem[\protect\citeauthoryear{{Tsupko} \& {Bisnovatyi-Kogan}}{{Tsupko} \&
  {Bisnovatyi-Kogan}}{2012}]{Tsupko2012OnGL}
{Tsupko} O.~Y.,  {Bisnovatyi-Kogan} G.~S.,  2012, \mn@doi [Gravitation and
  Cosmology] {10.1134/S0202289312020120}, \href
  {https://ui.adsabs.harvard.edu/abs/2012GrCo...18..117T} {18, 117}

\bibitem[\protect\citeauthoryear{{Tsupko} \& {Bisnovatyi-Kogan}}{{Tsupko} \&
  {Bisnovatyi-Kogan}}{2014}]{Tsupko2014GravitationalLI}
{Tsupko} O.~Y.,  {Bisnovatyi-Kogan} G.~S.,  2014, \mn@doi [Gravitation and
  Cosmology] {10.1134/S0202289314030153}, \href
  {https://ui.adsabs.harvard.edu/abs/2014GrCo...20..220T} {20, 220}

\bibitem[\protect\citeauthoryear{{Tuntsov}, {Walker}, {Koopmans}, {Bannister},
  {Stevens}, {Johnston}, {Reynolds}  \& {Bignall}}{{Tuntsov}
  et~al.}{2016}]{Tuntsov2015DYNAMICSM}
{Tuntsov} A.~V.,  {Walker} M.~A.,  {Koopmans} L. V.~E.,  {Bannister} K.~W.,
  {Stevens} J.,  {Johnston} S.,  {Reynolds} C.,   {Bignall} H.~E.,  2016,
  \mn@doi [\apj] {10.3847/0004-637X/817/2/176}, \href
  {https://ui.adsabs.harvard.edu/abs/2016ApJ...817..176T} {817, 176}

\bibitem[\protect\citeauthoryear{{Unzicker} \& {Case}}{{Unzicker} \&
  {Case}}{2005}]{Unzicker2005TranslationOE}
{Unzicker} A.,  {Case} T.,  2005, \mn@doi [arXiv e-prints]
  {10.48550/arXiv.physics/0503046}, \href
  {https://ui.adsabs.harvard.edu/abs/2005physics...3046U} {p. physics/0503046}

\bibitem[\protect\citeauthoryear{{Wang} \& {Mota}}{{Wang} \&
  {Mota}}{2020}]{Wang2020CanF}
{Wang} D.,  {Mota} D.,  2020, \mn@doi [\prd] {10.1103/PhysRevD.102.063530},
  \href {https://ui.adsabs.harvard.edu/abs/2020PhRvD.102f3530W} {102, 063530}

\bibitem[\protect\citeauthoryear{{Wei}, {Yang}  \& {Liu}}{{Wei}
  et~al.}{2015}]{Wei2014BlackHS}
{Wei} S.-W.,  {Yang} K.,   {Liu} Y.-X.,  2015, \mn@doi [European Physical
  Journal C] {10.1140/epjc/s10052-015-3469-7}, \href
  {https://ui.adsabs.harvard.edu/abs/2015EPJC...75..253W} {75, 253}

\bibitem[\protect\citeauthoryear{{Wong}, {Harko}, {Cheng}  \& {Gergely}}{{Wong}
  et~al.}{2012}]{2012PhRvD..86d4038W}
{Wong} K.~C.,  {Harko} T.,  {Cheng} K.~S.,   {Gergely} L.~{\'A}.,  2012,
  \mn@doi [\prd] {10.1103/PhysRevD.86.044038}, \href
  {https://ui.adsabs.harvard.edu/abs/2012PhRvD..86d4038W} {86, 044038}

\bibitem[\protect\citeauthoryear{{Wu}, {Li}, {Wu}  \& {Yu}}{{Wu}
  et~al.}{2014}]{Wu2014ConstrainsOF}
{Wu} J.,  {Li} Z.,  {Wu} P.,   {Yu} H.,  2014, \mn@doi [Science China Physics,
  Mechanics, and Astronomy] {10.1007/s11433-013-5302-3}, \href
  {https://ui.adsabs.harvard.edu/abs/2014SCPMA..57..988W} {57, 988}

\bibitem[\protect\citeauthoryear{{Wucknitz}, {Spitler}  \& {Pen}}{{Wucknitz}
  et~al.}{2021}]{Wucknitz2020CosmologyWG}
{Wucknitz} O.,  {Spitler} L.~G.,   {Pen} U.~L.,  2021, \mn@doi [\aap]
  {10.1051/0004-6361/202038248}, \href
  {https://ui.adsabs.harvard.edu/abs/2021A&A...645A..44W} {645, A44}

\bibitem[\protect\citeauthoryear{{Xie} \& {Deng}}{{Xie} \&
  {Deng}}{2013}]{Xie2013fG}
{Xie} Y.,  {Deng} X.-M.,  2013, \mn@doi [\mnras] {10.1093/mnras/stt991}, \href
  {https://ui.adsabs.harvard.edu/abs/2013MNRAS.433.3584X} {433, 3584}

\bibitem[\protect\citeauthoryear{{Yan}, {Zhang}, {Chen}, {Zhang}, {Cai}  \&
  {Saridakis}}{{Yan} et~al.}{2020}]{Yan2019InterpretingCT}
{Yan} S.-F.,  {Zhang} P.,  {Chen} J.-W.,  {Zhang} X.-Z.,  {Cai} Y.-F.,
  {Saridakis} E.~N.,  2020, \mn@doi [\prd] {10.1103/PhysRevD.101.121301}, \href
  {https://ui.adsabs.harvard.edu/abs/2020PhRvD.101l1301Y} {101, 121301}

\bibitem[\protect\citeauthoryear{{Yang}, {Hu}, {Cai}  \& {Wang}}{{Yang}
  et~al.}{2019}]{Yang2018NewPO}
{Yang} T.,  {Hu} B.,  {Cai} R.-G.,   {Wang} B.,  2019, \mn@doi [\apj]
  {10.3847/1538-4357/ab271e}, \href
  {https://ui.adsabs.harvard.edu/abs/2019ApJ...880...50Y} {880, 50}

\bibitem[\protect\citeauthoryear{{Zhang}, {Liguori}, {Bean}  \&
  {Dodelson}}{{Zhang} et~al.}{2007}]{Zhang2007ProbingGA}
{Zhang} P.,  {Liguori} M.,  {Bean} R.,   {Dodelson} S.,  2007, \mn@doi [\prl]
  {10.1103/PhysRevLett.99.141302}, \href
  {https://ui.adsabs.harvard.edu/abs/2007PhRvL..99n1302Z} {99, 141302}

\bibitem[\protect\citeauthoryear{{Zhao}, {Ren}, {Ilyas}, {Saridakis}  \&
  {Cai}}{{Zhao} et~al.}{2022}]{Zhao2022QuasinormalMO}
{Zhao} Y.,  {Ren} X.,  {Ilyas} A.,  {Saridakis} E.~N.,   {Cai} Y.-F.,  2022,
  \mn@doi [\jcap] {10.1088/1475-7516/2022/10/087}, \href
  {https://ui.adsabs.harvard.edu/abs/2022JCAP...10..087Z} {2022, 087}

\bibitem[\protect\citeauthoryear{{Zheng} \& {Huang}}{{Zheng} \&
  {Huang}}{2011}]{Zheng2010GrowthFI}
{Zheng} R.,  {Huang} Q.-G.,  2011, \mn@doi [\jcap]
  {10.1088/1475-7516/2011/03/002}, \href
  {https://ui.adsabs.harvard.edu/abs/2011JCAP...03..002Z} {2011, 002}

\bibitem[\protect\citeauthoryear{{Zitrin} \& {Eichler}}{{Zitrin} \&
  {Eichler}}{2018}]{Zitrin2018ObservingCP}
{Zitrin} A.,  {Eichler} D.,  2018, \mn@doi [\apj] {10.3847/1538-4357/aad6a2},
  \href {https://ui.adsabs.harvard.edu/abs/2018ApJ...866..101Z} {866, 101}

\makeatother
\end{thebibliography}

\appendix
\section{SIE lensing potential under \texorpdfstring{\bm{$f(T)$}} . gravity}
\label{SIE lensing potential under $f(T)$ gravity}
We simply present how we solve the lens equation to obtain the lens potential in $f(T)$.
$\Psi_{GR}$ stands for the gravitational potential of GR, while $\Psi_{\delta}$ represents the gravitational potential difference term between $f(T)$ and GR.
\begin{flalign}
&\Psi_{f(T)}\equiv\Psi_{GR} - \Psi_{\delta},\\
&\Psi_{GR}\equiv f(x)\tilde{\Psi}_{GR}(\varphi), \\
&\Psi_{\delta}\equiv g(x) \tilde{\Psi}_{\delta}(\varphi).
\end{flalign}
Thus the Poisson equations read
\begin{flalign}
&\frac{\partial^{2} \Psi_{GR}}{\partial x^{2}}+\frac{1}{x} \frac{\partial \Psi_{GR}}{\partial x}+\frac{1}{x^{2}} \frac{\partial^{2} \Psi_{GR}}{\partial \varphi^{2}}=\frac{\sqrt{f}}{x \Delta(\varphi) },\\
&\frac{\partial^{2} \Psi_{\delta}}{\partial x^{2}}+\frac{1}{x} \frac{\partial \Psi_{\delta}}{\partial x}+\frac{1}{x^{2}} \frac{\partial^{2} \Psi_{\delta}}{\partial \varphi^{2}}=\frac{A \sqrt{f}}{x^{3}\Delta^{3}(\varphi)}.
\end{flalign}
Since $x$ and $\varphi$ are not coupled, we have
\begin{flalign}
&\tilde{\Psi}_{GR}(\varphi)+\frac{d^{2}}{d \varphi^{2}} \tilde{\Psi}_{GR}(\varphi)=\frac{\sqrt{f}}{\Delta(\varphi)},\\
&\tilde{\Psi}_{\delta}(\varphi)+\frac{d^{2}}{d \varphi^{2}} \tilde{\Psi}_{\delta}(\varphi)=\frac{A \sqrt{f}}{\Delta^{3}(\varphi)}.
\end{flalign}
We already know the solution for $\tilde{\Psi}_{GR}$ is
\begin{align}
f(x) &= x,\nonumber \\
\tilde{\Psi}_{GR}(\varphi)&= \frac{\sqrt{f}}{f^{\prime}}\left[\sin \varphi \arcsin \left(f^{\prime} \sin \varphi\right)+\cos \varphi \operatorname{arcsinh}\left(\frac{f^{\prime}}{f} \cos \varphi\right)\right],\nonumber \\
\tilde{\Psi}_{GR}(x,\varphi) &= x \frac{\sqrt{f}}{f^{\prime}}\left[\sin \varphi \arcsin \left(f^{\prime} \sin \varphi\right)+\cos \varphi \operatorname{arcsinh}\left(\frac{f^{\prime}}{f} \cos \varphi\right)\right].
\end{align}
Same as $\tilde{\Psi}_{GR}$, we used Green's function method for $\tilde{\Psi}_{\delta}$,
\begin{align}
\tilde{\Psi}_{\delta}(\varphi)&= \int_{-\infty}^{\infty} d \varphi^{\prime} \hat{G}\left(\varphi, \varphi^{\prime}\right) \mathcal{J}(\varphi^{\prime}),\\
\text{with}\quad\mathcal{J} &= \frac{A \sqrt{f}}{\Delta^{3}(\varphi)},
\end{align}
and Green's function can be constructed as
\begin{flalign}
\hat{G}\left(\varphi, \varphi^{\prime}\right)=\frac{1}{W} & {\left[\theta\left(\varphi-\varphi^{\prime}\right) \hat{\phi}_{-}\left(\varphi^{\prime}\right) \hat{\phi}_{+}(\varphi)\right.} \\
+&\left.\theta\left(\varphi^{\prime}-\varphi\right) \hat{\phi}_{-}(\varphi) \hat{\phi}_{+}\left(\varphi^{\prime}\right)\right].
\end{flalign}
$\theta(x)$ is the step function, $W$ is the Wronskian of $\hat{\phi}_{+}$ and $\hat{\phi}_{-}$,
\be
W \equiv \hat{\phi}_{-}(\varphi) \partial_{\varphi} \hat{\phi}_{+}(\varphi)-\hat{\phi}_{+}(\varphi) \partial_{\varphi} \hat{\phi}_{-}(\varphi).
\ee
Combining the two equations above, the solution for $\tilde{\Psi}_{\delta}$ is
\begin{flalign}
\tilde{\Psi}_{\delta}(\varphi)=& \frac{1}{W} \hat{\phi}_{+}(\varphi) \int_{0}^{\varphi} d\varphi^{\prime} \hat{\phi}_{-}\left(\varphi^{\prime}\right) \mathcal{J}\left(\varphi^{\prime}\right) \\
&+\frac{1}{W} \hat{\phi}_{-}(\varphi) \int_{\varphi}^{2\pi} d\varphi^{\prime} \hat{\phi}_{+}\left(\varphi^{\prime}\right) \mathcal{J}\left(\varphi^{\prime}\right),
\end{flalign}
For homogenoeous equation of $\tilde{\Psi}_{2}$, we write
\begin{align}
&\hat{\phi}_{+} = \sin \varphi,
\quad\hat{\phi_{-}}=\cos \varphi,\\
&W =\sin ^{2} \varphi+\cos ^{2} \varphi=1, \\
&\hat{G} = \theta(\varphi-\varphi^{\prime}) \cos\varphi^{\prime} \sin\varphi+\theta(\varphi^{\prime}-\varphi) \cos\varphi\sin\varphi^{\prime}.
\end{align}
Thus,
\begin{flalign}
\tilde{\Psi}_{\delta} &= \sin \varphi \int_{0}^{\varphi} \cos \varphi^{\prime} \mathcal{J}^{\prime} d \varphi^{\prime}+\cos \varphi \int_{\varphi}^{2 \pi} \sin \varphi^{\prime}  \mathcal{J}^{\prime} d \varphi^{\prime}\nonumber\\
&=A \sqrt{f}\left[\frac{\sin ^{2} \varphi}{\Delta(\varphi)}+\frac{1}{f^{2}} \frac{\cos ^{2} \varphi}{\Delta(\varphi)}\right].
\end{flalign}
Taking the solution back to the original Poisson equation, $g(x) = \frac{1}{x}$,
\begin{flalign}
\Psi_{f(T)} &=\Psi_{GR}-\Psi_{\delta} \nonumber \\
&=f(x) \tilde{\Psi}_{GR}(\varphi)-g(x) \tilde{\Psi}_{\delta}(\varphi) \nonumber \\
&=x \cdot \frac{\sqrt{f}}{f^{\prime}}\left[\sin \varphi \arcsin \left(f^{\prime} \sin \varphi\right)+\cos \varphi \operatorname{arcsinh}\left(\frac{f^{\prime}}{f} \cos \varphi\right)\right] \nonumber \\
&-\frac{1}{x} \cdot \frac{A\sqrt{f}}{f^2} \cdot  \Delta(\varphi).
\end{flalign}





\label{lastpage}
\end{document}